\pdfoutput=1 
\documentclass{aa}  

\usepackage{graphicx}
\usepackage{txfonts}
\PassOptionsToPackage{hyphens}{url}\usepackage{hyperref}

\newcommand{\teff}{\rm T_{eff}}
\newcommand{\logg}{\log(g)}
\newcommand{\feh}{\rm [Fe/H]}
\newcommand{\mh}{\rm [M/H]}

\newcommand{\apoteff}{\textsc{TEFF}}
\newcommand{\apologg}{\textsc{LOGG}}
\newcommand{\apofeh}{\textsc{FE\_H}}
\newcommand{\apomh}{\textsc{M\_H}}
\newcommand{\apomhuncal}{\textsc{M\_H}_{\rm unc.}}
\newcommand{\apoalpha}{\textsc{ALPHA\_M}}
\newcommand{\apoparam}{\textsc{PARAM}}
\newcommand{\apofparam}{\textsc{FPARAM}}
\newcommand{\starbad}{\textsc{STAR\_BAD}}
\newcommand{\chibad}{\textsc{CHI2\_BAD}}
\newcommand{\metalsbad}{\textsc{METALS\_BAD}}
\newcommand{\metalswarn}{\textsc{METALS\_WARN}}

\begin{document}

   \title{The metal-poor tail of the APOGEE survey}
   \titlerunning{The metal-poor tail of APOGEE I}

   \subtitle{I. Uncovering $\feh < -2.5$  stars from the inner Galaxy to the Magellanic Clouds }

   \author{M. Montelius
          \inst{1}\fnmsep\thanks{\email{montelius@astro.rug.nl}}
		  \and
          E. Starkenburg\inst{1}
          \and
                    H. C. Woudenberg\inst{1}
          \and
                    A. Angrilli Muglia\inst{1}
          \and
                    A. Ardern-Arentsen\inst{2}
          \and
                    A. Viswanathan\inst{1,3}
        \and
            A. Bystr{\"o}m\inst{4}
        \and 
            A. Helmi\inst{1}
        \and
            N. Martin\inst{5,6}
        \and
            T. Matsuno\inst{7}
        \and 
            C. Navarrete\inst{8}
        \and
            J. Navarro\inst{9}        
 }

\institute{Kapteyn Astronomical Institute, University of Groningen, Landleven 12, NL-9747 AD Groningen, the Netherlands     
          \and
             Institute of Astronomy, University of Cambridge, Madingley Road, Cambridge CB3 0HA, UK 
        \and
              Dept. of Physics and Astronomy, University of Victoria, P.O. Box 3055, STN CSC, Victoria BC V8W 3P6, Canada 
        \and 
            Institute for Astronomy, University of Edinburgh, Royal Observatory, Blackford Hill, Edinburgh EH9 3HJ, UK 
        \and 
            Université de Strasbourg, CNRS, Observatoire astronomique de Strasbourg, UMR 7550, F-67000 Strasbourg, France
        \and
            Max-Planck-Institut für Astronomie, Königstuhl 17, D-69117 Heidelberg, Germany 
        \and
            Astronomisches Rechen-Institut, Zentrum für Astronomie der Universität Heidelberg, Mönchhofstraße 12-14, 69120 Heidelberg,
Germany 
         \and
            Université Côte d’Azur, Observatoire de la Côte d’Azur, CNRS, Laboratoire
Lagrange, Bd de l’Observatoire, CS 34229, 06304, Nice Cedex 4, France 
        \and
            CIfAR Senior Fellow and Professor. Department of Physics and Astronomy, University of Victoria, Victoria, BC, Canada V8P 5C2
            }

   \date{Received 1 July 2025 / Accepted 26 September 2025 }

  \abstract
   {The most metal-poor stars in our Galaxy contain important clues of its earliest history, particularly those occupying the inner regions of the Galaxy. In the search for such metal-poor stars, large spectroscopic surveys are an invaluable tool. However, the spectra of metal-poor stars can be difficult to analyse because of the relative lack of available lines, which can also lead to misclassification. } 
   {We aim to identify the stars observed by the APOGEE survey that are below the metallicity limit of APOGEE's analysis. For the highest confidence candidates, we classify the orbital properties of the stars to investigate whether their orbital distribution matches what we would expect for stars that are this metal poor and to select especially interesting targets for spectroscopic follow-up purposes.
   }
   {We examined the properties derived by APOGEE for metal-poor stars from the literature to find signatures of stars with a metallicity below the range of the grid used for spectral analysis. Once identified within APOGEE, we computed the orbits of our metal-poor candidates using AGAMA.
   }
   {The calibrated APOGEE stellar parameters provide a clear signature of the most metal-poor stars in the survey, indicated by null values for their metallicities while having effective temperatures and surface gravities determined by the pipeline. From comparison with the literature, we find that, within a temperature range of 3700 - 6700 K and above a threshold of S/N > 30, the vast majority of APOGEE stars without calibrated metallicities are very metal poor. Additional cleaning by visual inspection of the spectra improved the purity of the sample further. The radial velocities provided by APOGEE, Gaia DR3 positions and astrometry along with spectrophotometric distances derived in this work allowed for the computation of their orbits. In this work, we carefully selected 289 very metal-poor red giant stars (likely below \feh = -2.5) from the APOGEE results. Our sample contains 16 very metal-poor member candidates of the Magellanic Clouds, 14 very metal-poor stars with orbits confined to the inner Galaxy, and 13 inner Galaxy halo interlopers. These samples significantly add to the very metal-poor stars known in these regions and allow for a more detailed picture of early chemical evolution across different environments. }
   {}


   \maketitle

\section{Introduction}
The chemical abundances in a star's photosphere are largely unchanged from their birth until the final stages before their death. The preserved chemical record of the oldest stars therefore presents an important tool in the study of the formation and evolution of our Galaxy, and we refer to this type of study as Galactic Archaeology \citep{FreemanBland-Hawthorn2002}. Within a closed environment, the metallicity of stars shows a strong correlation with their age. Cosmological simulations suggest that the majority of very and extremely metal poor (VMP and EMP, $\feh$ < $-2$ and $-3$, respectively\footnote{Metallicities in this paper are defined as [A/B] = $\log (N_\mathrm{A}/N_\mathrm{B})_* - \log (N_\mathrm{A}/N_\mathrm{B})_\odot$, with $N_\mathrm{A}$ and $N_\mathrm{B}$ being the number densities of elements A and B. }, \citealt{BeersChristlieb2005}) stars in the Galaxy today formed before redshift $z = 2.5$ \citep{El-Badry2018}. However, in a galaxy such as the Milky Way with accretion of both gas and other galaxies, the model of a closed system breaks down and with it the relation between metallicity and age. Moreover, the pace of chemical enrichment varies in different Galactic environments. As the inner parts of the Milky Way are thought to have formed early and at relatively high, the VMP stars residing here are expected to be among the oldest stellar populations, even if they are not the most metal-poor stars \citep[e.g.][]{WhiteSpringel2000,Zolotov2009,Tumlinson2010,Starkenburg2017simulations,El-Badry2018}.
 
 Identifying stars this metal poor has been successfully done with (narrow-band) photometry \citep[e.g.][]{Beers1985, Christlieb2002, Keller2007Skymapper, Starkenburg2017PristineI, Cenarro2019JPLUS} and large spectroscopic surveys, including Gaia-ESO \citep{GaiaESO2012}, the Large sky Area Multi-Object fiber Spectroscopic Telescope \citep[LAMOST\footnote{\url{http://www.lamost.org/public/?locale=en}},][]{LAMOST2012}, GALactic Archaeology with HERMES \citep[GALAH,][]{GALAH2015}, Apache Point Observatory Galactic Evolution Experiment \citep[APOGEE][]{APOGEE2017}, and Gaia RVS \citep{RecioBlanco2023GaiaRVS}. These surveys gather millions of spectra, and upcoming projects such as WEAVE \citep{WEAVE2012}, 4MOST \citep{4MOST2019overview}, MOONS \citep{MOONS2014}, and DESI \citep{DESI2013,DESI2025stellarDR1} are expected to observe millions more spectra in the coming years. An inherent difficulty in identifying metal-poor stars from medium-resolution surveys of this kind is that as stars become more metal poor, their spectral lines become weaker and harder to analyse. Due to the computational cost involved in traditional spectral analysis, the work of finding and analysing these stars is therefore often left to additional studies that make use of more specialised methods of analysis than what was used for the majority of observed spectra \citep[see e.g.][]{topos2013, SEGUE_trashcan2016, Li2022, Hughes2022GALAH-EMP-tSNE, LAMOST_trashcan2023, Matsuno2022, Viswanathan2024a}. 

 The APOGEE survey is different from most other stellar surveys in terms of the wavelength region observed. APOGEE observed in the near-infrared H-band, between 15140 $\AA$ and 16960 $\AA$. Observing in the near-infrared gives a unique advantage regarding the Galactic regions that can be studied, as infrared light is less affected by dust extinction than optical light. This enabled APOGEE to observe stars in obscured parts of the Galactic disc and bulge, in addition to the Galactic halo, where most metal-poor stars are found. Dedicated searches in the inner Galaxy have already confirmed the existence of a VMP population with orbits that confine them to the bulge region \citep[see e.g.][]{Howes2014Gaia-ESO-bulge, LuceyCOMBS2019, Arentsen2020PIGS-I, Rix2022, ArdernArentsen2024}, although the samples are still small, particularly in the most dust-obscured regions. As with the bulge, the disc of our Galaxy also hosts mostly more metal-rich populations, but recent work has shown that there are sizeable populations of stars down to the ultra metal-poor regime of [Fe/H] $< -4$ that have disc-like orbits \citep[][]{Sestito2019}.
 
While sizeable samples of EMP stars exist in the current literature \citep[see for an overview the SAGA or JINA databases,][]{SAGAIV2017, Abohalima2018}, a vast majority of these reside in the (local) Milky Way halo, therefore limiting the exploration of early chemical evolution across different environments. A homogeneous analysis of EMP stars across the Milky Way bulge, disc, and halo -  with different star formation histories and varying paces of chemical enrichment - will be very valuable for our understanding of early enrichment. The APOGEE footprint is particularly promising in this regard, as it includes all of these different Milky Way components. Moreover, APOGEE also includes pointings towards many of the known Milky Way satellites, including the Small and Large Magellanic Clouds \citep[e.g.][]{Hasselquist2021ApJ...923..172H}. These large dwarf galaxies present an opportunity to study early chemical evolution in a different environment, where the current studies of the lowest metallicity population are very limited in number \citep{Reggiani2021AJ....162..229R, Oh2023Magellanic, Chiti2024NatAs...8..637C, Oh2024, Akshara2024dist}.

To derive stellar parameters quickly and accurately, the APOGEE Stellar Parameter and Chemical Abundance Pipeline \citep[ASPCAP;][]{ASPCAP2016} uses the FERRE optimisation code \citep{FERRE2006}. The method uses a grid of synthetic spectra to compare to the observed spectra, with a range of metallicities down to $\feh$ = $-2.5$ (which means that lower metallicities cannot be measured by APOGEE).
 Several studies have been conducted to analyse larger samples of APOGEE spectra \citep[see e.g.][]{Szabolcs2015, Hawkins2016}, and there are more specialised studies looking for chemically enhanced stars \citep{Kielty2017cemp-apogee, Fernandez-Trincado2020aluminium-enhanced}. As the lines in the H-band are generally weaker than in the optical, this limits the analysis of the most metal-poor stars. As a result, the APOGEE survey has not been re-analysed to search for stars below APOGEE's metallicity floor as of yet. 

 In this paper, we describe a method to identify a pure sample of metal-poor stars beyond the ASPCAP metallicity lower bound of $\feh$ = $-2.5$. The main purpose of identifying such a sample is to search for the oldest stars in our Galaxy in the dust-obscured regions where optical studies struggle. Additionally, we identify VMP stars in other regions of the sky targeted by APOGEE, among them the Magellanic Clouds. While in this work we focus on identifying such stars, validating their selection, and investigating their orbits in the Galaxy, a companion paper (Montelius, Angrilli Muglia et al., in prep.) analyses the remaining lines in their spectra and discusses their magnesium and silicon abundances, as well as their metallicities. We will refer to this companion paper as paper II in this work.
 
 This work is structured as follows. In Sect. \ref{sec:method}, we describe the method used to select metal-poor stars and how contamination was minimised in the sample. Section \ref{sec:comp} subsequently validates the selection with a literature comparison. We review the dynamics of our clean sample in Sect. \ref{sec:dyn} and present the resulting samples of stars in the inner Galaxy and in the Magellanic Clouds. In Sect. \ref{sec:summ}, we discuss future prospects of using this sample and how future H-band surveys can include such stars in their analysis. 

\section{Sample selection method}
\label{sec:method}
We started our selection of metal-poor candidates in APOGEE from the latest and most complete APOGEE catalogue, DR17 \citep{APOGEEDR17}. For stars with multiple observations, we only kept the highest signal-to-noise ratio (S/N) spectra for each unique APOGEE ID. While APOGEE provides a wide range of quality flags that can be used to select reliable targets, we opted to use an external catalogue to remove contamination. 

We first cross-matched our APOGEE targets with Gaia DR3 \citep{GaiaDR32022} using the EDR3 Gaia source ID provided by APOGEE. To remove possible galaxies and quasars from the sample, we required a probability of $\rm P_{star}>0.95$ from the Gaia Discreet Source Classifier \citep{Delchambre2023DSC}. 11020 sources are removed with this cut. 

The spectra gathered by the APOGEE survey have all been analysed by ASPCAP to derive stellar parameters and abundances. ASPCAP uses the optimisation code FERRE \citep{FERRE2006} to derive stellar parameters by comparing observed spectra to a grid of synthetic spectra computed with Synspec \citep{hubney2021arXiv210402829H}. The grid is populated to a lowest metallicity boundary of $\mh$ = $-2.5$, and since ASPCAP does not extrapolate outside the grid, lower metallicities cannot be measured.
The results of the ASPCAP analysis have been calibrated to better align with independent measurements. The direct FERRE measurements are referred to as the `uncalibrated' (in line with their tags in the $\apofparam$ parameters file from APOGEE) while the final products are called  `calibrated' (following the tag $\apoparam$) in this work.
The main spectroscopic parameters derived by ASPCAP of interest in this paper are effective temperature ($\apoteff$), surface gravity ($\apologg$), overall metallicity ($\apomh$) and $\feh$ ($\apofeh$)\footnote{While often used interchangeably, $\apofeh$ is measured only from Fe lines, while $\apomh$ is measured from the whole spectrum.}. 
For DR17, the details of how parameters are calibrated by ASPCAP will be provided in Holtzman et al. (in prep.), but a summary is provided on the SDSS website\footnote{\url{https://www.sdss4.org/dr17/irspec/aspcap/}} and the procedure used in DR16 is described in \cite{Jonsson2020ASPCAP}. Uncalibrated parameters are used to derive the metallicity and elemental abundances, and calibration is only performed on $\apoteff$, $\apologg$, and $\apoalpha$. In the transfer from the $\apoparam$ to the so called `named tags', ASPCAP will not assign a value if the star has either the $\starbad$ or $\chibad$ flags\footnote{See \url{https://www.sdss4.org/dr17/irspec/apogee-bitmasks/}}. In previous data releases the flag $\starbad$ has been assigned if a star has any parameters close to the edge of the grid, but in DR17 this is not the case for the metallicity in particular. For metal-poor stars at the edge of the metallicity grid, this means that there is a chance that they will get assigned $\apoteff$ and $\apologg$, but not $\apomh$, if the overall analysis is deemed successful, but the metallicity is out of range for the grid, or close to the grid edge. 

\begin{figure}
\includegraphics[width=0.5\textwidth]{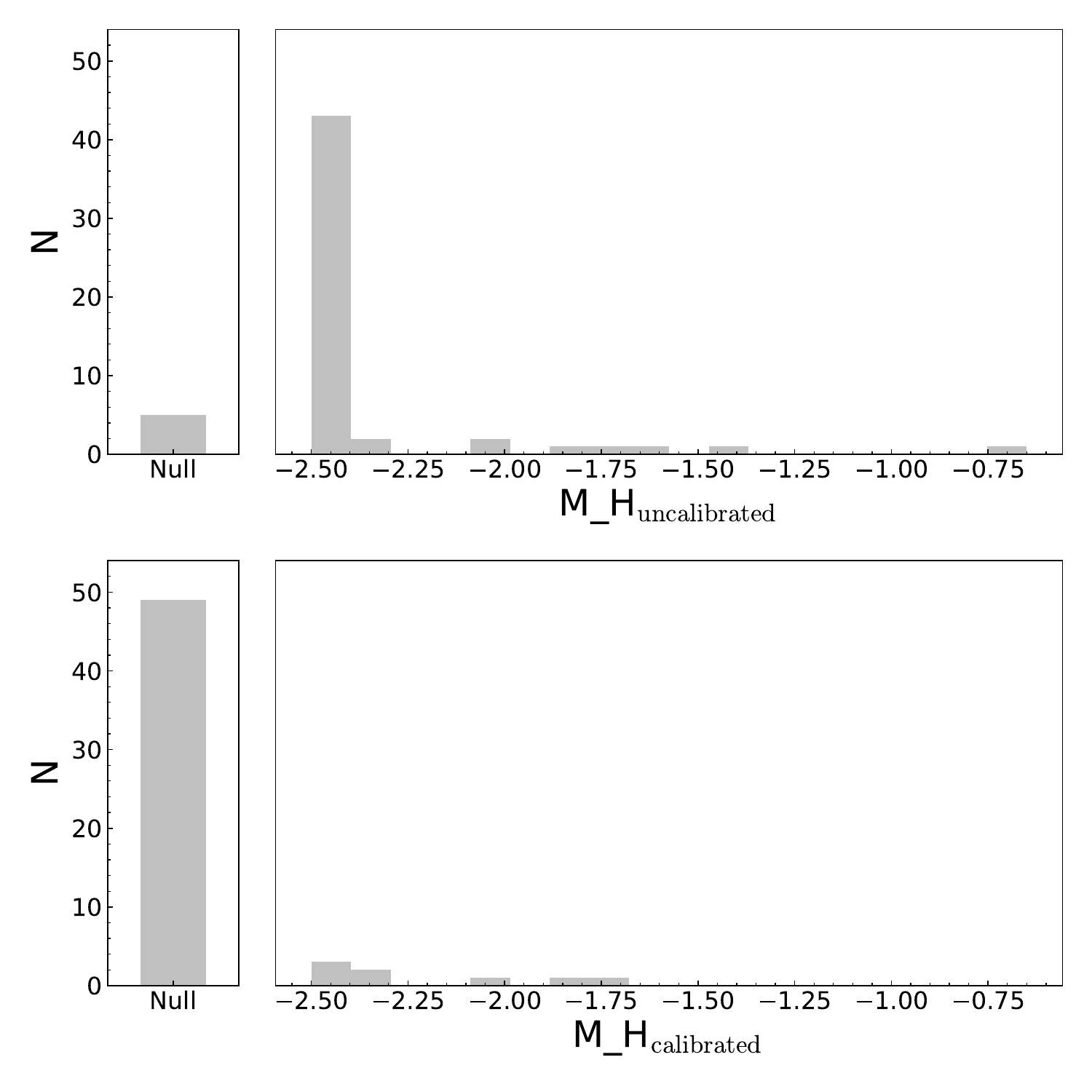}
\caption{Metallicity distribution functions for SAGA stars with literature $\feh<-2.5$. Top panel:  uncalibrated metallicities from APOGEE. Bottom panel: calibrated APOGEE metallicities. The number of null values for each metallicity is shown with the same y-axis.}
    \label{fig:SAGA_MDF}
\end{figure}

To find out what happens to the spectra of stars below the grid edge of $\apomh$ = $-2.5$ when they go through ASPCAP, we selected a sample of stars with literature $\feh$ < $-2.5$. A total of 57 stars matching this criteria were taken from the SAGA database \citep[][see Sect. \ref{sec:SAGA}]{SAGAIV2017}. 
 Figure \ref{fig:SAGA_MDF} shows the metallicity distribution function (MDF) for these reference stars, with the upper panel showing uncalibrated metallicities ($\apomhuncal$) and the lower panel showing calibrated values ($\apomh$). Both MDFs also include the null values where no value has been assigned. 
 
Five of the stars get null values for both metallicities. These all have S/N < 5. The majority of the stars are close to the edge of the grid in the upper panel, with 43 stars that have $\apomhuncal$ < $-2.4$. Only a very small number of the stars get significantly higher values, with 5 stars $\apomhuncal$ >$-2$. 
A similar result can be seen in the lower panel of Figure \ref{fig:SAGA_MDF} that shows the calibrated values. The majority of the stars, 49 of them, get null values. 
When a value of $\apomh$ is assigned, we note that on average this is 0.45 dex higher than the metallicities from SAGA for the same stars. 

\begin{figure}
\includegraphics[width=0.5\textwidth]{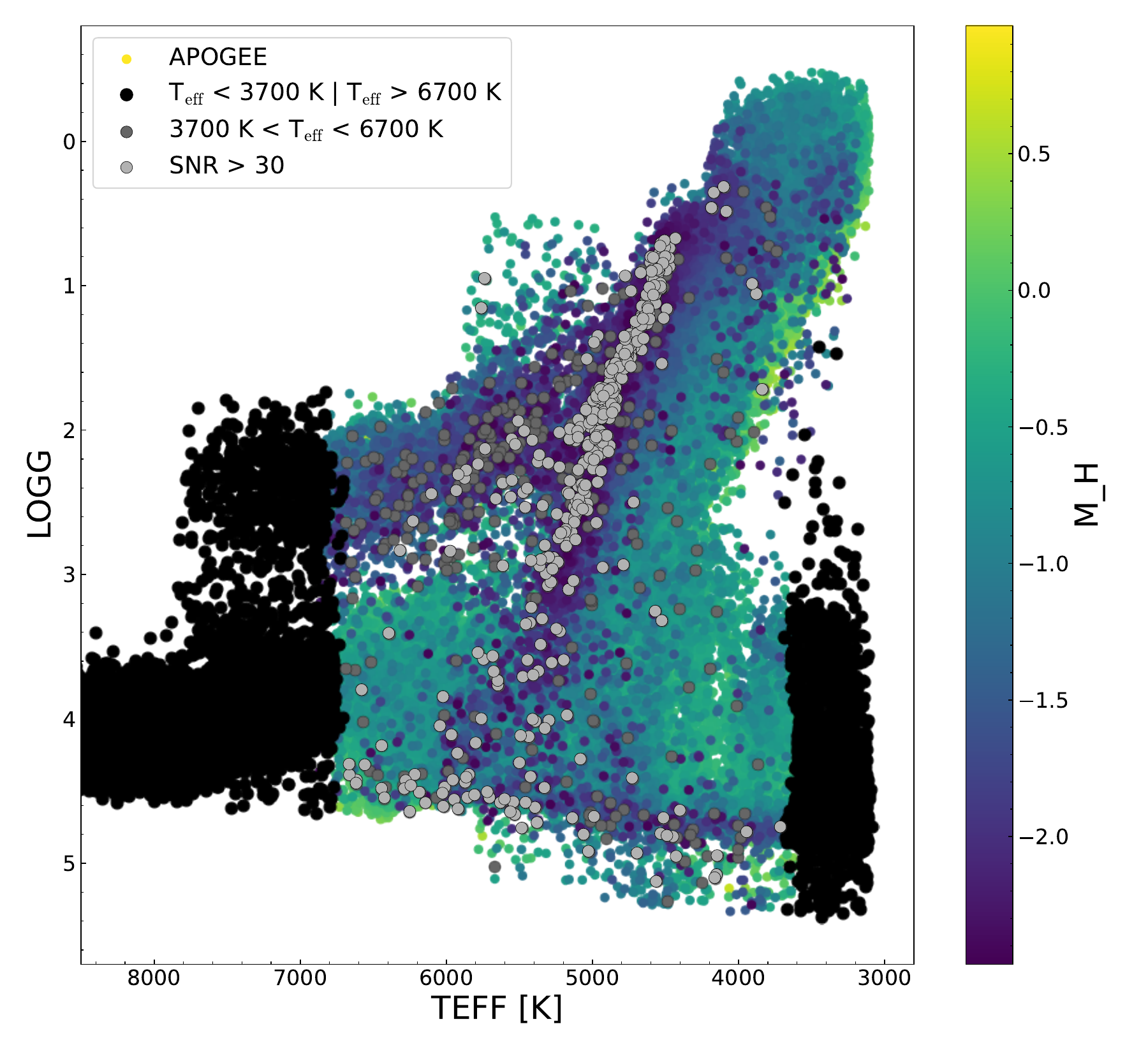}
\caption{Kiel diagram of all APOGEE stars, colour-coded by their calibrated $\apomh$ values, plotted with the lowest metallicity stars on top. The stars with null metallicity measurements are plotted in greyscale: stars with $\apoteff$ too high or too low to reliably get metallicity measurements are shown in black; stars with S/N < 30 are shown in dark grey; stars with S/N > 30 are shown in light grey.}
\label{fig:Kiel_selection_M_H}
\end{figure}

To better understand the properties of such stars in the larger APOGEE catalogue, we show a Kiel diagram of all stars with calibrated $\apoteff$ and $\apologg$ but either no calibrated $\apomh$ or no $\apofeh$ in Figure \ref{fig:Kiel_selection_M_H}. To give context for where metal-poor stars are usually found in the Kiel diagram, the APOGEE stars with calibrated $\apomh$ are shown colour-coded by their metallicity and with the most metal poor stars plotted on top. The stars without $\apomh$ have been split into three categories
\begin{enumerate}
\item Stars with $\apoteff$ that is either lower than 3700 K or higher than 6700 K are shown in black. 
\item Spectra outside the temperature range in 1, with signal to noise ratio (S/N\footnote{We use the S/N reported by APOGEE, which is calculated per pixel.}) lower than 30, are shown in dark grey.
\item Spectra outside the temperature range in 1, with S/N higher than 30 are shown in light grey. We refer to this sample as our candidates for the purpose of this work.
\end{enumerate}

\begin{figure}
\includegraphics[width=0.5\textwidth]{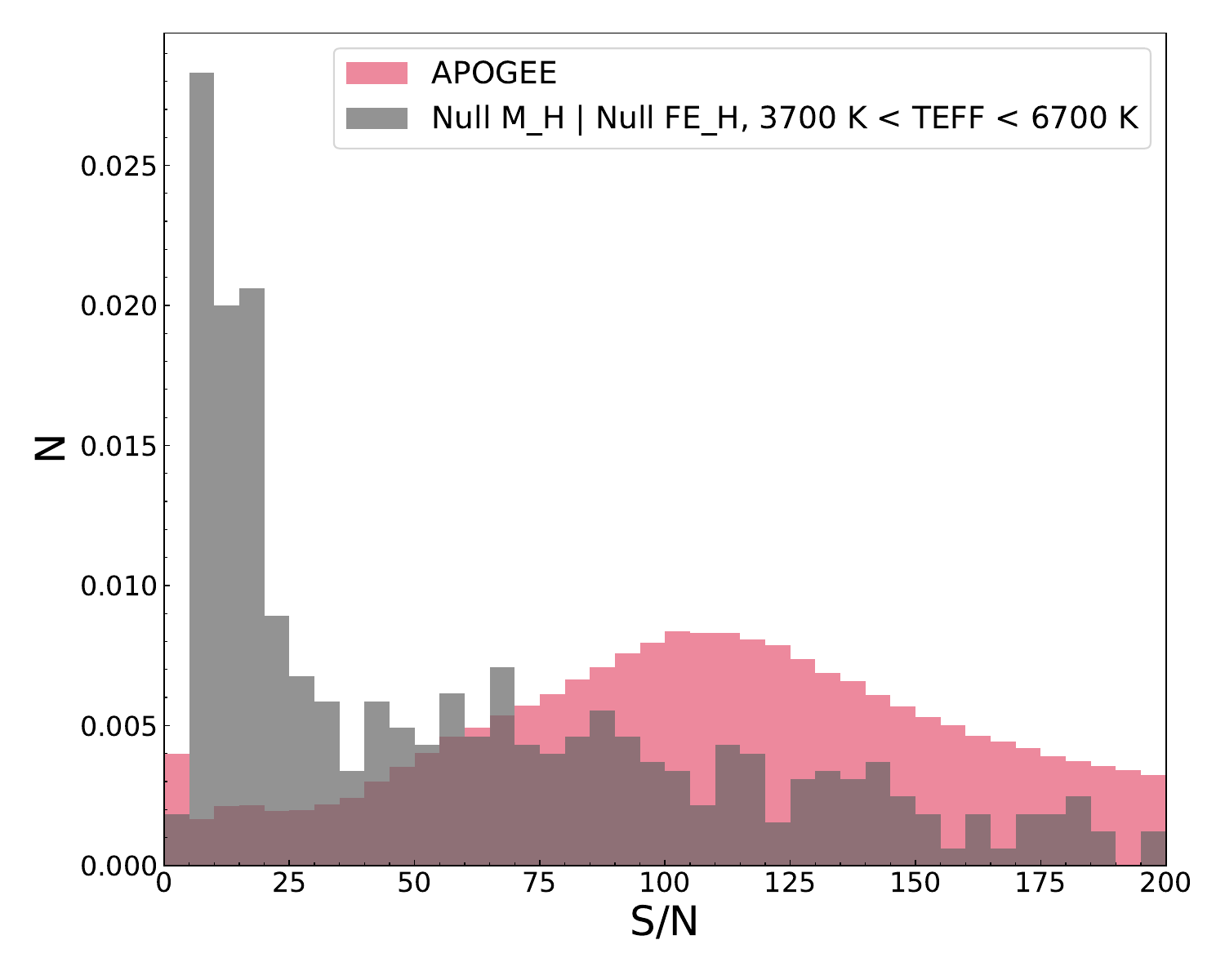}
\caption{Normalised histograms of the S/N for all APOGEE stars (shown in red) and for the stars with calibrated $\apoteff$ but without calibrated metallicity (shown in grey). For very low S/N values the stars without metallicity are clearly overrepresented compared to the overall distribution. At S/N $\gtrsim$ 30 the distribution flattens, suggesting that the reason for failing to derive a calibrated metallicity is not low S/N. }
\label{fig:SNR_hist}
\end{figure}

\begin{figure*}
    \centering
    \includegraphics[width=\linewidth]{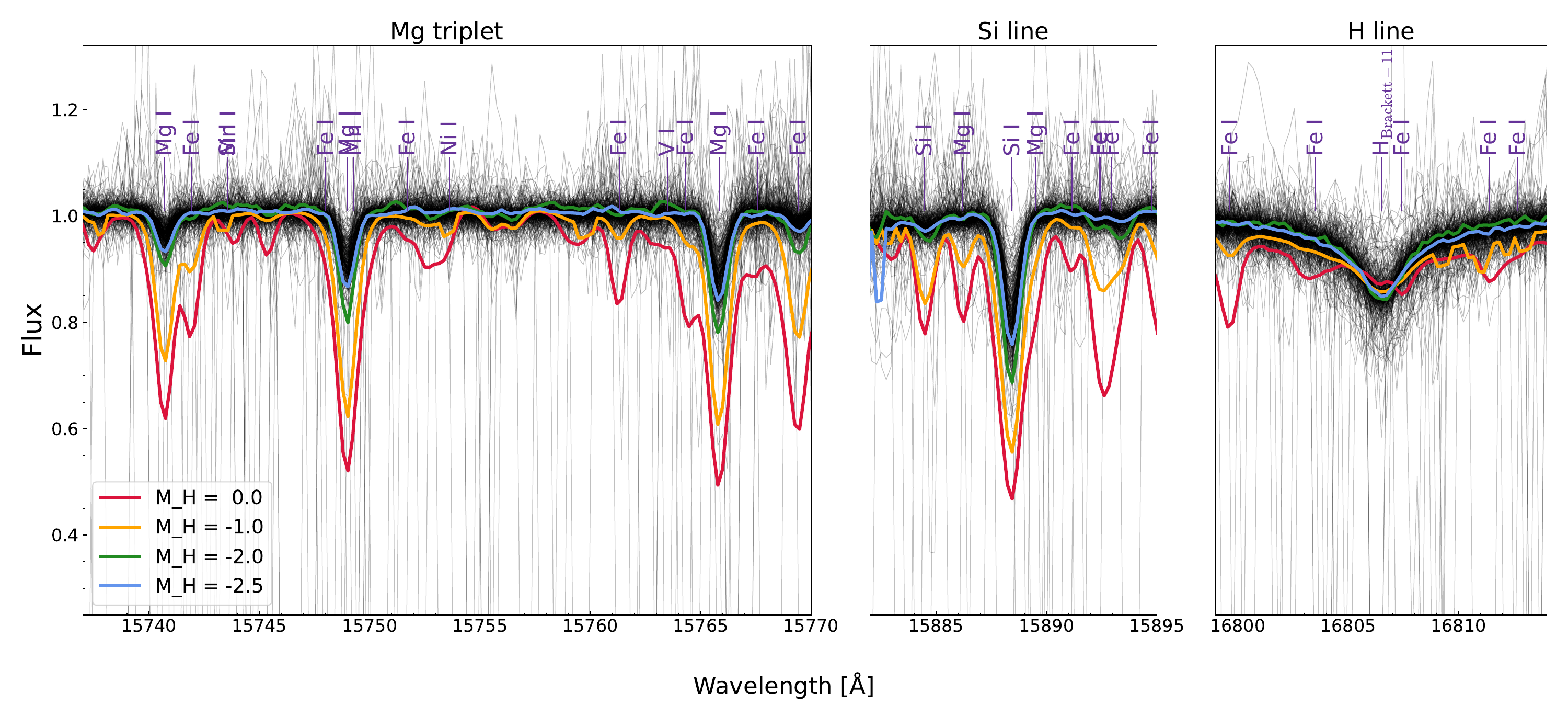}
    \caption{Spectral view of APOGEE spectra for the 326 candidate VMP stars in this work (grey lines), after cleaning the sample. Vertical thin lines across the wavelength space connect to artefacts in some of these spectra that are not affecting the overall analysis. Also shown are the averaged APOGEE spectra for red giants with temperatures and gravities in a range comparable to that of the bulk of our sample (4100 K $< \teff <$ 5200 K and 0.5 $< \logg <$ 3.5) and a distinct range in metallicities. The comparison clearly shows that our selected sample stars are indeed very metal poor.}
    \label{fig:MgSiH_spectra}
\end{figure*}

Separating out the temperature range of 3700 K < $\apoteff$ < 6700 K was motivated by the sharp drop in the number of stars with calibrated metallicities outside of that range. For the hotter stars, the fraction of stars with calibrated metallicities drops from $99.4\%$ in the range 6600 K - 6700 K, to $84.9\%$ and $10.6\%$ for 6700 K - 6800 K and 6800 K - 6900 K, respectively. For the cooler stars The giants and dwarfs behave differently, with $99.8\%$ of stars with $\apoteff$ < 3800 K and $\apologg$ < 3 getting a value for $\apomh$. For the main sequence stars the picture is different, with the percentage of stars with $\apomh$ and $\apologg$ > 3, dropping from $99.9\%$, to $89.8\%$, to $2.8\%$ for the ranges 3700 K - 3800 K, 3600 K - 3700 K and 3500 K - 3600 K.
This rapid change clearly indicates that APOGEE is unable to reliably constrain the metallicity of stars with these temperatures, potentially due to the weak metal lines in the hot stars and the strong molecular lines for the cool stars. As such, we discarded these stars as potential metal-poor candidates.
Of the remaining stars with moderate temperatures, a large portion of the candidate stars appear to line up well in the Kiel diagram with the location of the most metal poor of the giants with calibrated $\apomh$. This indicates that they are probably genuine metal-poor stars, and not contamination. The decision to split the sample at S/N=30 (used as the threshold for dark and light grey points in Figure \ref{fig:Kiel_selection_M_H}) is informed by Figure \ref{fig:SNR_hist} showing normalised histograms for both the entire APOGEE sample and for stars with calibrated $\apoteff$ between 3700 K and 6700 K and no calibrated metallicities. There is a clear overabundance of stars with low S/N, plausibly interpreted as spectra where the low S/N is the cause of the missing calibrated metallicity. At higher S/N values, the distribution follows the overall S/N distribution of APOGEE more closely, suggesting that S/N is not the main factor for why the spectra have no calibrated $\apomh$. Instead, we hypothesise it is their true very metal-poor nature that causes the inability of the pipeline to provide a calibrated $\apomh$.

As a final cleaning step in our selection procedure, we visually checked all APOGEE spectra that pass the above-mentioned criteria (i.e., S/N $> 30$, no calibrated metallicities, and TEFF between 3700 and 6700 K). In this check, we removed stars where the Mg I triplet at 15740 \AA~ or the Si I line at 15888 \AA~ were un-analysable. This removed $101$ stars, a large fraction of which are identified as emission-line stars or spectra disrupted by edge effects from the gap between spectral regions close to 15800 \AA. This left us with a sample of 326 candidate VMP stars. Figure \ref{fig:MgSiH_spectra} shows an overview of all these APOGEE spectra as thin grey lines. The coloured thick lines in this figure are averaged APOGEE spectra for red giants (of similar stellar parameters as the majority of our sample) in different metallicity ranges. A comparison between these coloured lines and the grey lines clearly shows that these stars are indeed very metal poor. A majority of them agree best with the red giant branch spectra for $\mh = -2.5$, further validating our sample selection. We investigate this sample further in the next sections, while we refer to paper II for a quantitative spectral analysis of the Mg and Si lines shown in Figure \ref{fig:MgSiH_spectra}.

\section{Literature comparison}
\label{sec:comp}
To build intuition about the stars selected from APOGEE, we compare to other surveys that can reliably measure metallicities for VMP and EMP stars. These comparison samples are briefly described below, a description of the cross-matched results with our selected metal-poor candidates is presented in Sect. \ref{sec:results}.

\subsection{The Pristine-Gaia data release}
\label{sec:pristine}
With the third data release from the Gaia survey \citep{GaiaDR32022} came the first release of Gaia XP spectrophotometry for 219 million stars \citep{DeAngeli2023}. \cite{PristineXXIII2023} has used this dataset to derive photometric metallicities by constructing a synthetic narrow band filter over the Ca II H and K lines, mimicking the filter used in the Pristine survey \citep{Starkenburg2017PristineI}. We note that several other studies have also successfully derived metallicities from Gaia XP spectra using a variety of techniques \citep[e.g.][]{Andrae2023a,Andrae2023b,Bellazzini2023,Zhang2023,Xylakis2024}, but as the Pristine Gaia-synthetic metallicities from \cite{PristineXXIII2023} are especially calibrated for metal-poor stars we prefer them for the purpose of our work. We additionally note that \cite{PristineXXVI2024} have demonstrated that the Pristine-Gaia metallicities are highly effective at identifying stars with $\feh$ < $-2.5$. In their spectroscopic follow-up study of stars with photometric metallicities below $-2.5$, $76$\% of the spectroscopic metallicities satisfy this criteria and only $3$\% are outliers with $\feh$ > $-2$ \citep{PristineXXVI2024}.

\subsection{SAGA}
\label{sec:SAGA}
The SAGA database \citep{SAGAIV2017} collects high-resolution spectroscopic measurements, mostly of metal-poor stars in the Milky Way and local dwarf galaxies. For this paper we have used their recommended data set, version April 7, 2021. It contains only one record per star, prioritising sources as described in \cite{SAGAIV2017}. As $\feh$ is not available for all stars, values have been selected according to the priority list: 1: $\feh$, 2: ([Fe I/H] + [Fe II/H])/2, 3: [Fe I/H], 4: [M/H]. We refer to this combination of abundances as $\feh$ for simplicity.

\subsection{LAMOST}

With over 11 million publicly released low-resolution spectra to date, the LAMOST survey \citep{LAMOST2015} is one of the largest spectroscopic surveys. As their Low-Resolution Spectroscopic Survey (LRS) mode uses a resolution of R$\approx$1800 with a multi-fibre telescope, it can target a wide range of sources. This wide net is ideal for gauging the level of contamination from both high-metallicity stars and other sources. 
As the LAMOST analysis pipeline also has a grid edge at $\feh=-2.5$, we use metallicities obtained by re-analysing LAMOST DR10 spectra with FERRE \citep{FERRE2006}, following the procedure described in Sect. 3.1 of \cite{AnkeLAMOST2025}. Stars with LAMOST spectra with S/N < 10, as measured by FERRE, are excluded to ensure the measurements are of sufficient quality. We will refer to this sample as LRS - FERRE in this work.

\subsection{DESI MWS DR1}
While the Dark Energy Spectroscopic Instrument (DESI) Survey \citep{DESI2013} is primarily focused on measuring redshifts for external galaxies, there is also a large stellar survey component, called the Milky Way Survey (MWS). In the first public data release \citep{DESIDR1}, 4 million stars have been observed. Similar to LAMOST, the relatively low resolution (R$\approx$2000-5200) and large sample size makes DESI a good choice to validate the level of contamination. Additionally, the stellar DESI pipelines have the capability to measure metallicities to [Fe/H] = $-4$ and as the survey footprint focuses on the Galactic halo, many VMP and EMP stars are included. 
For our purposes, we use the parameters from DESI's RV pipeline, requiring RR\_SPECTYPE == STAR, RVS\_WARN == 0, SN\_R > 10, SUCCESS == 1 and VSINI < 30.

\subsection{Gaia RVS re-analysis}
\label{sec:RVS}

A final source for very low-metallicity candidates that we considered in this study are results from the Gaia RVS spectra \citep{RecioBlanco2023GaiaRVS}. Similar to the other datasets listed above, we focused in this work on analyses of these spectra that specialise in the VMP regime. Specifically, we used the dataset of \cite{Matsuno2022}, who use additional photometric information to provide more precise metallicity estimates for VMP stars analysed by the GSP-Spec module. The additional information reduces the degeneracy between temperature and metallicity.

We also used the dataset from \cite{Viswanathan2024a}, who selected VMP stars using literature photometric metallicities and subsequently (re-)analysed the Gaia RVS spectra, adding $\sim$ 750 VMP stars that previously got either no metallicity estimate from the RVS pipeline, or were flagged as unreliable. We combined this with results from the spectroscopic follow up of metal poor-candidates from the Pristine survey by \cite{PristineXXVI2024}, which analyses spectra from the Isaac Newton Telescope/Intermediate Dispersion Spectrograph (INT/IDS) in the Calcium triplet region with the same method used in \cite{Viswanathan2024a}.

\begin{figure}
    \centering
    \includegraphics[width=0.5\textwidth]{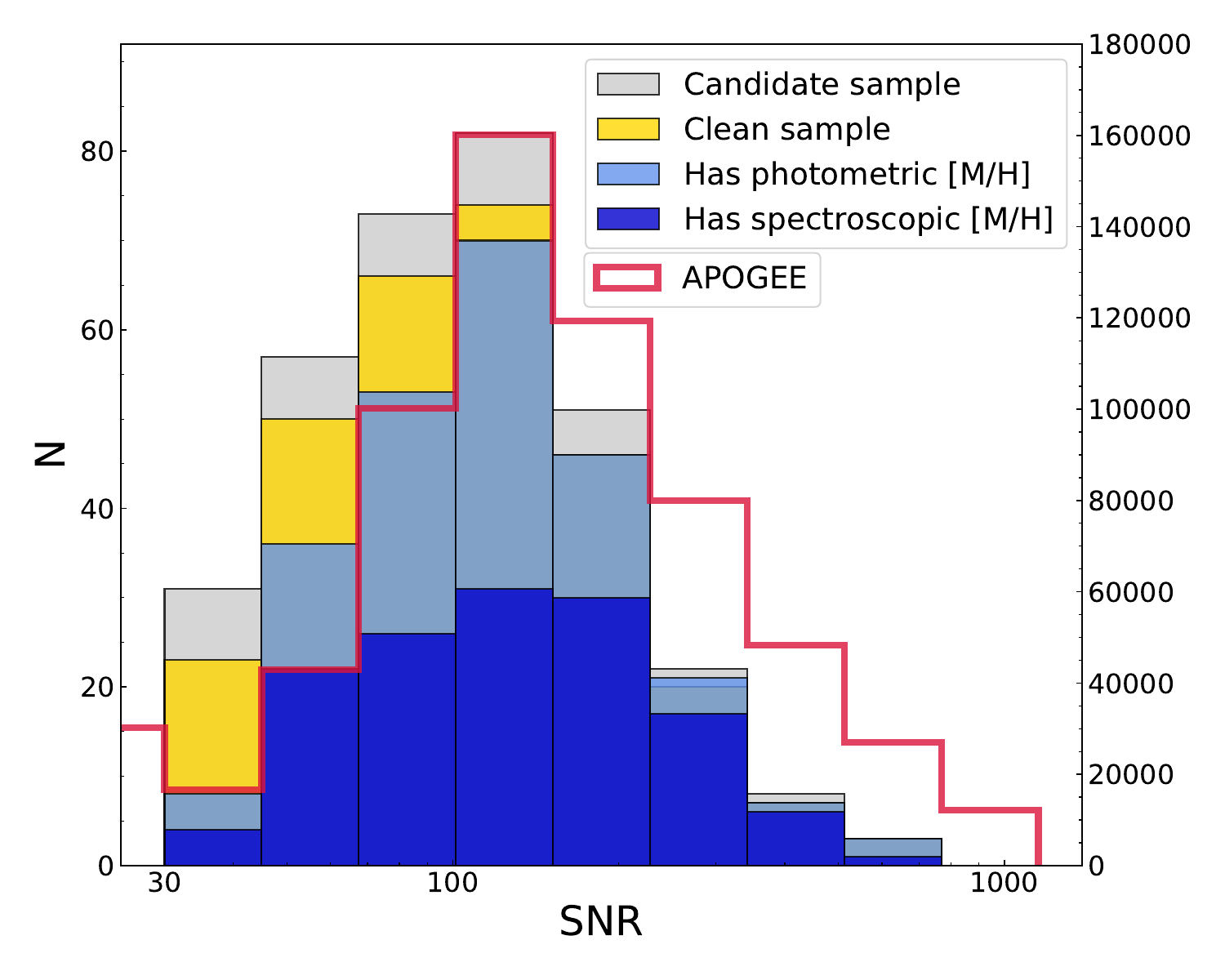}
    \caption{Normalised histogram of the S/N on a log scale of the candidate stars, shown in grey, and of the subset of candidates with values from the literature sources in Sect. \ref{sec:comp}, including the additional sources in section \ref{sec:more_lit}, split into photometric and spectroscopic metallicities. The overall S/N distribution of APOGEE is shown as a red outline, with the right side y-axis tracking their numbers. }
    \label{fig:all_gtnmonhgsnr_SNR}
\end{figure}

\begin{figure*}[!h]
    \centering
    \includegraphics[width=0.9\textwidth]{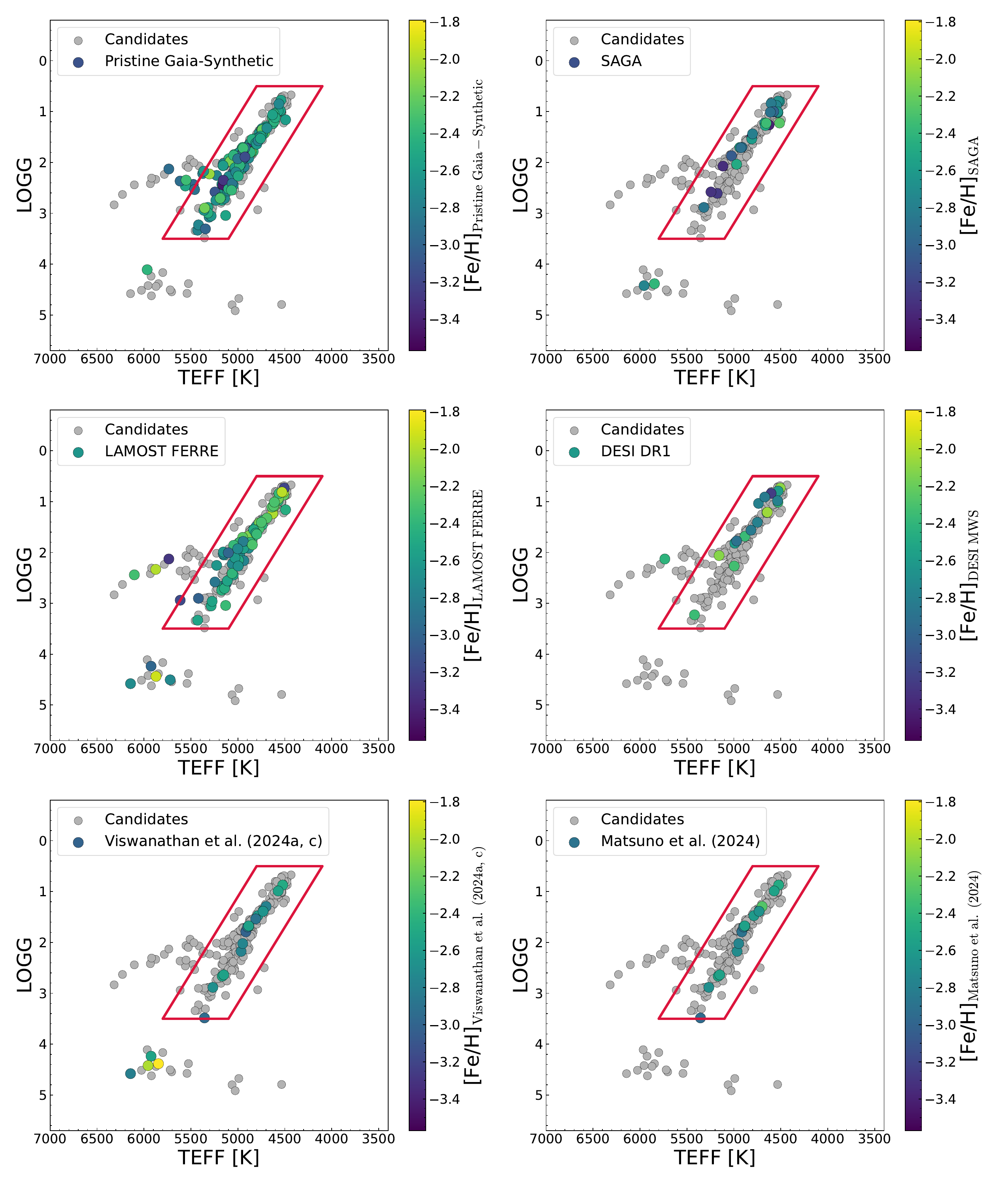}
    \caption{Kiel diagrams of the candidate metal-poor stars. The candidates with S/N > 30 have been cross-matched with the various reference values presented in Section \ref{sec:comp}. To aid comparison across the samples, a common range in colours has been assigned. The red outline shows the refined selection of candidates on the metal-poor red giant branch where comparison data has been sufficient to check for contamination.}
    \label{fig:Kiel_hex}
\end{figure*}

\subsection{Additional literature sources}
\label{sec:more_lit}
While the literature sources described above contain the largest samples of overlap, we note that some additional sources contain a few stars from our sample. These stars are included in the following sections under the name `previous literature' samples and listed here for completeness. In particular, we note that some stars have photometric metallicities in the (recommended) Table 2 of \cite{Andrae2023b}, while they do not get a photometric metallicity in the \citet{PristineXXIII2023} data release (even though both datasets are based on the same Gaia XP spectra, they use different quality cuts that account for these differences). Some additional spectroscopic measurements for (very small) samples of overlapping stars have been taken from GALAH DR4 \citep{2025PASA...42...51B}, LAMOST MRS \citep{LAMOST2012}, the Pristine Inner Galaxy Survey (PIGS) spectroscopic follow-up \citep{PIGSII2020MNRAS.496.4964A, ArdernArentsen2024}, follow-up of SEGUE EMP candidates \citep{SEGUE_trashcan2016} and follow-up of Pristine VMP candidates \citep{Aguado2019MNRAS.490.2241A}. 
Finally, crossmatches to the SIMBAD Astronomical Database yielded 17 additional stars with metallicities from different studies \citep{simbad2000AJ....120.1351S, simbad2012PASP..124..519F, simbad2014ApJ...784..170X, simbad2014ApJ...794...60T, simbad2015MNRAS.448.2717W, simbad2018AJ....156..257S, simbad2020A&A...641A.127R, simbad2020ApJ...894...34D, simbad2021ApJ...913...11L, simbad2022ApJ...926...26S, simbad2022MNRAS.513.3993O, simbad2023MNRAS.524..577O}. 

\subsection{Comparison results}
\label{sec:results}

To evaluate the level of contamination in the candidate sample, we cross-matched it with the literature sources described in Sections \ref{sec:pristine} - \ref{sec:more_lit}. 

As the S/N of the spectra is important for a star to get a metallicity, Figure \ref{fig:all_gtnmonhgsnr_SNR} shows the normalised histogram of all candidate stars and of the ones with some literature value, along with the overall S/N distribution of the full APOGEE sample. The candidates without literature values appear to be overrepresented at lower S/N values. This is as expected and most likely a consequence of the lower S/N spectra belonging to stars that are more difficult to observe, both fainter and in more dust obscured regions. As the literature studies are all based on optical light they are more heavily affected by dust extinction and are therefore less likely to observe these stars.

As Figure \ref{fig:Kiel_selection_M_H} shows that the candidate sample covers a large part of the Kiel diagram, it is important to check how the literature values are distributed in this space as well. Figure \ref{fig:Kiel_hex} shows the Kiel diagrams for the candidates and the crossmatched literature samples. The majority of the literature sources are concentrated on the metal-poor giant branch, where most of the candidates lie as well. The rest of the Kiel diagram is significantly less covered by the literature sources, and most of the stars with literature values above $\feh$ = $-2$ in metallicity are there.
Due to the lack of sufficient testing data and the increased contamination, we selected a clean sample of stars by defining a region around the metal-poor giant branch, marked as a red outline. While the stars outside of this region still appear to be metal poor, we caution that the level of contamination is likely higher.
Our resulting `clean' sample consists of 289 stars, 53 of which are not at all in the literature compilations mentioned above, and 83 of which only have a photometric metallicity (so also no radial velocity measurement) in the literature. The remaining 38 stars outside the clean selection have 16 stars with literature metallicities. The lack of coverage of literature values in the Kiel diagram for these remaining candidates leads us to focus on the clean sample for this paper, as we do not know how representative the available literature values are. The spectra of both samples will be analysed in paper II. 

\begin{figure}
\centering
\includegraphics[width=0.5\textwidth]{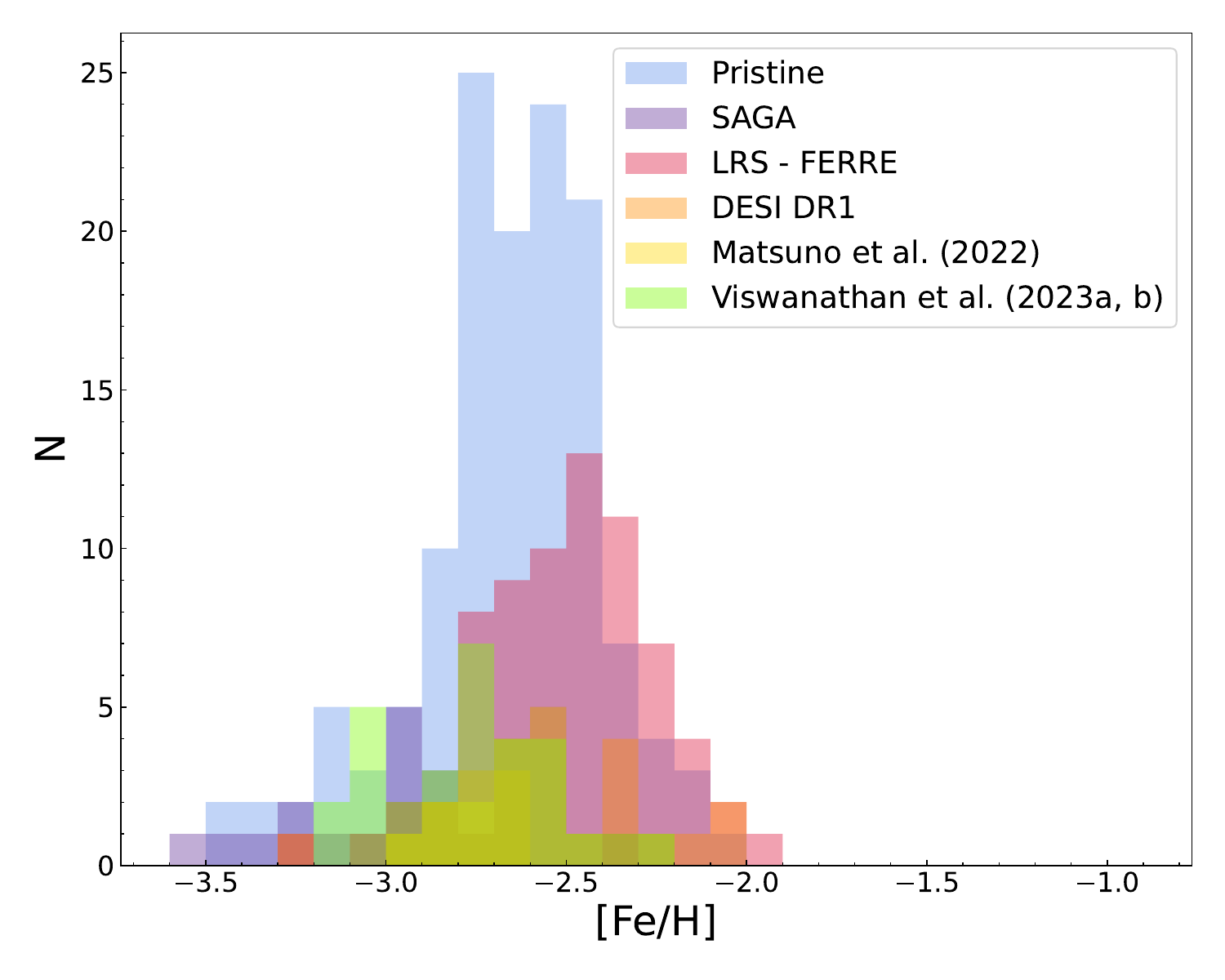}
\caption{Metallicity distribution function for all candidate stars with the metallicities from the literature sources in Sect. \ref{sec:comp}. }
    \label{fig:all_gtnmonhgsnr_MDF}
\end{figure}

Figure \ref{fig:all_gtnmonhgsnr_MDF} shows the resulting MDF with the literature metallicities for the resulting cleaned sample taking into account the Kiel diagram selection. The resulting sample is indeed very clean, with 99.7\% of the literature metallicities for these red giant branch (RGB) stars below $-2.0$, while $70$ \% have at least one measurement below the grid edge at $-2.5$\footnote{We note that, as indicated in the previous subsections, not all literature comparisons cover the full metallicity range as some only measure $\feh > -2.5$ (e.g. LAMOST), or are conversely only focused on the VMP regime \citep[e.g.][]{SAGAIV2017, Li2022, Viswanathan2024a}. Nevertheless, we have included samples that do include the full metallicity range.}.  

\section{Dynamics of APOGEE VMP stars}
\label{sec:dyn}

In this era of large spectroscopic datasets, APOGEE remains unique in its capability to look through dusty regions of the Galaxy, resulting in a more complete coverage of all Galactic components compared to other spectroscopic surveys in the optical. Although 
APOGEE cannot measure metallicities for these VMP stars
because of the vanishing Fe lines, the mere existence of APOGEE spectra for the sample of 289 (highly likely) VMP red giant branch stars constructed above do add to sparse literature samples in dust-obscured regions. From the APOGEE spectra radial velocity information for these stars can be derived. This important dynamical information, when combined with Gaia-measured proper motions and a distance estimate for the stars, allows for the reconstruction of their orbits.

In this section, we carry out an orbital analysis for our clean sample selected above. In particular, we focused on determining whether any of the stars in our sample orbit in the inner Galaxy, where infrared spectroscopy's advantage of lower extinction is most relevant. Additionally, we have taken advantage of APOGEE's special pointing strategy for dwarf galaxies \citep[e.g.][]{Hasselquist2021ApJ...923..172H} and we isolate VMP stars in the Magellanic Clouds, another stellar population where limited information on early chemical evolution is available in the current literature.

\subsection{Distances}
\label{sec:dist}
While a necessary first step to access 3D distributions and orbits of stars, determining distances to stars throughout the Galaxy is a difficult task. In particular, this is true towards the inner Galaxy, as uncertainties on parallax measurements are relatively large at this distance, and, additionally, the dense and complex dust environment makes dust-correction to isochrone fitting particularly challenging. In this work, we derived distances by fitting the spectroscopic calibrated stellar parameters TEFF and LOGG from APOGEE to BaSTI isochrones \citep{Hidalgo2018BaSTI}. The BaSTI isochrone set used has an alpha-enhancement of 0.4 \citep{BaSTIalpha2021}, and has been shown to perform better for fitting distances with stellar parameters \citep{Akshara2024dist} when compared to MIST \citep{MIST2016ApJS..222....8D} and PARSEC \citep{PARSEC2012MNRAS.427..127B} isochrones. The preference for the BaSTI isochrones is attributed in part by \cite{Akshara2024dist} to the alpha enhancement of the model, as the PARSEC models share this feature and they are in better agreement than the MIST models which lack this important abundance variation. In our testing, the fit of the BaSTI models to the stellar parameters was better than the fit to the MIST and PARSEC models. The APOGEE $\apoteff$ values have shown offsets to the $\teff$ in all tested isochrones, most significantly for turn-off stars, an issue that has also been seen in other surveys \citep{daSilva2023A&A...677A..74D}. As we limit our investigation to the clean sample of giant branch stars, the effect of the offset is minimised. For our comparison, we set the age to 10 Gyr and assume a metallicity of $\feh$ = $-2.5$. We note that, even though our stars might have a lower metallicity than this value, this will affect the isochrones minimally as VMP isochrones cluster very closely together on the red giant branch. The \textit{Gaia} data was dereddened using the Total Galactic Extinction (TGE) module \citep{TGE2023, dustmaps2018} from \textit{Gaia} with the temperature-dependent conversion to extinction for the \textit{Gaia} G band from \citet{photutils2022}.

To make sure that $\logg$ and $\teff$ are both taken into account when determining the nearest point in two dimensions to the isochrone, we multiplied $\logg$ values by 1000 for $\logg$ <= 2 and 3000 for $\logg$ > 2. This procedure is very similar to the procedure applied in \cite{Akshara2024dist}, but the exact factors for scaling the $\logg$ are not the same, motivated by the greater accuracy of the APOGEE $\logg$ compared to the values derived from Gaia XP used in \cite{Akshara2024dist}. Fitting a spline to $\teff$ and the modified $\logg$ and finding the closest point on the line for all stars results in absolute magnitudes that can be compared to the apparent \textit{Gaia} G magnitudes. As a test sample for validation of our method, we selected stars with high quality parallax distances, $\frac{\pi}{\Delta\pi}>5$, between parallax distances $4.3$ kpc $< r_{\pi}< 11.3$ kpc to match the distance of stars with Galactocentric radii $R_{\rm GC} < 3.5$ kpc, yielding us 100 stars. The parallax distances were computed by inverting \textit{Gaia} DR3 parallaxes corrected with the zero point offset from \cite{Gaiazeropoint2020}. When comparing our distances with the parallax-based distances, the validation sample has a mean and standard deviation of $\mu = -0.09$ kpc and $\sigma = 1.71$ kpc. To get an estimate of the uncertainty in our results, we took the standard deviation of the difference between our distance and the parallax distance, divided by the parallax distance $\frac{\Delta r}{r_{parallax}} \sim0.25$, and we therefore adopt 25 $\%$ as our distance uncertainty.

To further evaluate the quality of our distances we compare our results to those from the \textit{StarHorse} code \citep{Starhorse2023}. \textit{StarHorse} incorporates spectroscopic, photometric and astrometric data into a Bayesian framework to derive distances along with other stellar parameters. Approximately 73\% of our sample has distances derived by \textit{StarHorse}. We note that for the parallax-based distance validation sample defined above, the \textit{StarHorse} perform slightly worse compared to our method, with $\mu = -0.58$ kpc and $\sigma = 1.89$ kpc. More details on the comparison of these samples are presented in Appendix \ref{app:dists}.

While these tests are constrained to stars within 11.3 kpc, a comparison of a very similar method at larger distances carried out by \citet{Akshara2024dist} indicated the scatter in distance estimates for the Galactic halo is 20\% for good stars with good-quality stellar parameters and up to 40\% when the stellar parameters are less good quality. We note that this implies that we can expect to have a 1-$\sigma$ range for distances towards the Magellanic Clouds, for instance, that is of order 10 - 25 kpc.

\subsection{Orbit integration }
\label{sec:orbit_integration}
To investigate the dynamical properties of the stars selected with this method, we have used \texttt{AGAMA} \citep{Vasiliev2019AGAMA}. Because of our focus on the inner Galaxy, we included the impact of a rotating bar on our results by adopting the barred Galactic potential model from \cite{Hunter2024} with the \cite{Sormani2022} bar model. 
We assumed a pattern speed of $\Omega_b = -37.5$~km s$^{-1}$ kpc$^{-1}$ and a bar angle of $\alpha = -25$~degrees, consistent with recent literature estimates \citep{2025NewAR.10001721H}. 
Our coordinate transformation to Galactocentric Cartesian and cylindrical coordinates followed the assumptions made in \cite{Dodd2023}. 
The solar velocities (U=11.1 km/s, V=12.24 km/s, W=7.25 km/s) were taken from \cite{Schonrich2010}, the velocity of the local standard of rest is (229.8 km/s) from \cite{Sormani2022} and the solar Galactocentric radius (8.2 kpc) is from \cite{McMillan2017}.
The orbits were integrated for $\sim$10 Gyr. 
For the orbital analysis, we limited the orbits we study to bound orbits, with E < -40000 km$^2$s$^{-2}$ at present time. This leaves out seven stars. An exception to this cut in orbital energy was made for stars in dwarf galaxies. As our potential only models the Milky Way, orbital information for stars that are bound to any satellite galaxies will not be accurate. Because of this, these stars are marked in all plots of orbital properties. Dwarf galaxy stars have been selected by comparison with \cite{Battaglia2022}, finding a total of 34 stars. Additionally, as detailed further in the following subsection, 16 other stars could be identified as (likely) members of the Magellanic Clouds \citep[not included in][]{Battaglia2022} and these are also marked separately.

\subsection{Galactic distribution}
\label{sec:distribution}
\begin{figure}
\centering
\includegraphics[width=0.5\textwidth]{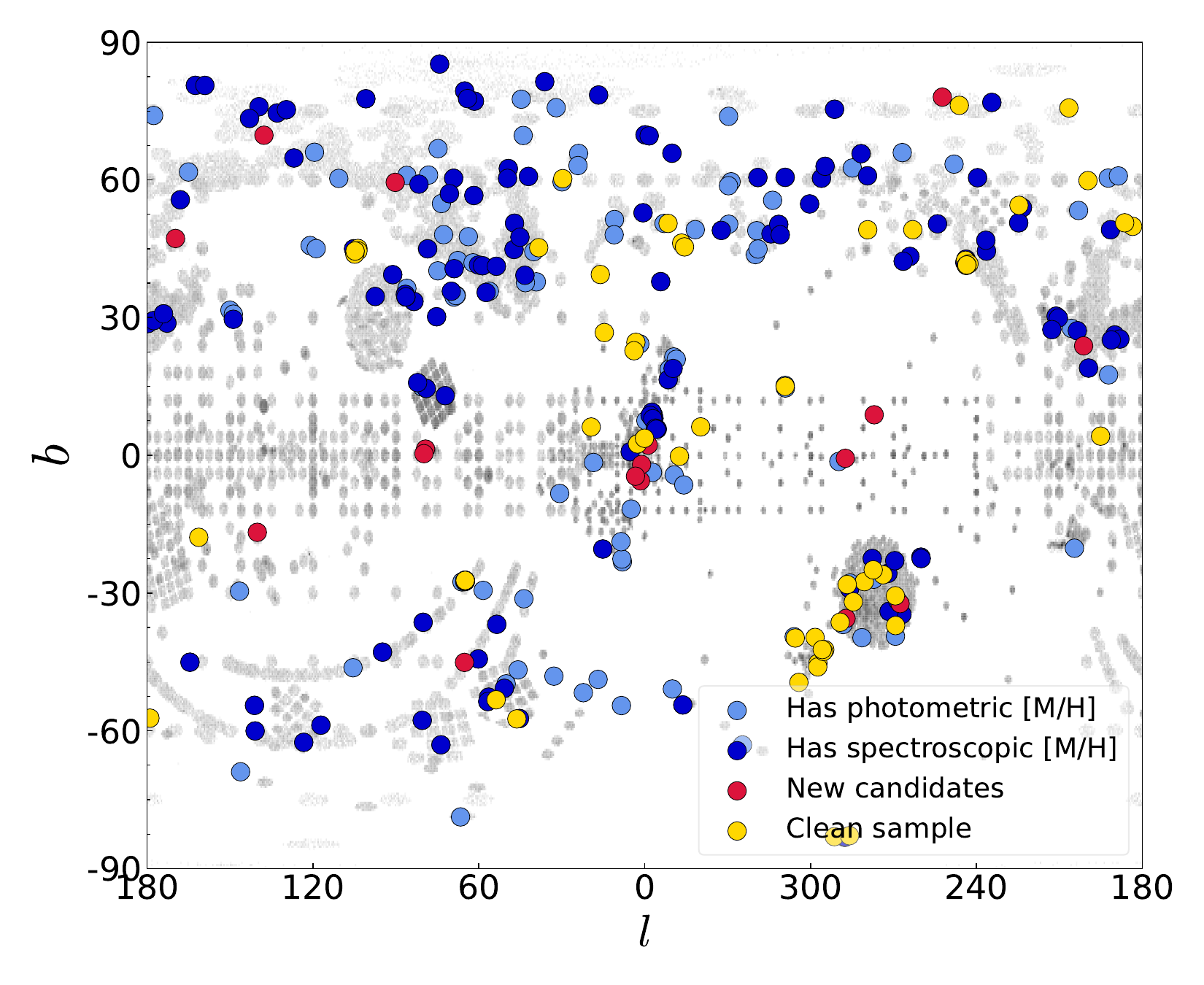}
\caption{Position on the sky in galactic coordinates the clean sample of RGB stars highlighted in yellow and the lower confidence sample of main sequence and horizontal branch sample shown in red. The candidates with literature metallicities described in Sect. \ref{sec:comp}, including the additional sources in Sect. \ref{sec:more_lit}, are shown in blue, and the entire APOGEE sample is in grey. }
    \label{fig:Skycov}
\end{figure}

The distribution of all candidates on the sky are shown in Figure \ref{fig:Skycov}. In total, 139 have literature spectroscopic metallicities, 106 stars have only photometric metallicities, and 82 are new metal-poor candidates, 62 of which belong to the high confidence RGB sample. The candidates are distributed throughout the APOGEE footprint, with 73\% above |$b$| = $20^{\circ}$, corresponding to halo stars that are known to be the most metal-poor population, as well as stars in dwarf galaxies. The remaining $27$\% that are seen closer to the disc are concentrated around the inner $20^{\circ}$, where we expect metal-poor stars in the bulge.

\begin{figure}
    \centering
    \includegraphics[width=\linewidth]{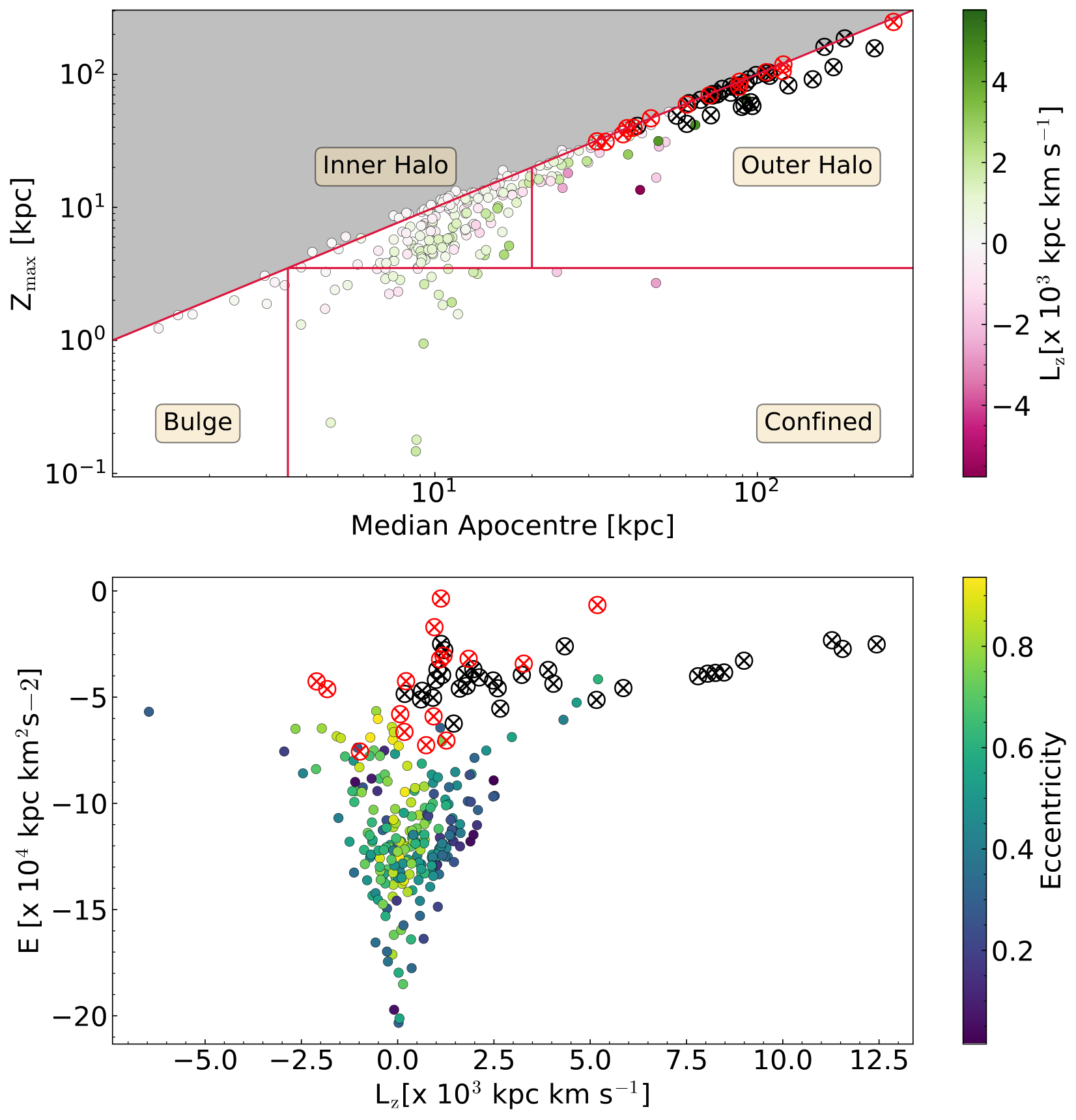}
    \caption{Above: Dynamical groupings following \cite{PristineXXVI2024}. We note that we used the median apocentres, which (due to the propagated uncertainties in the orbit integration) causes a bit of scatter around the diagonal line. 
    Below: Energy versus the z component of the angular momentum. Likely members of a dwarf galaxies based on classifications from \cite{Battaglia2022} have been marked with crossed out circles (red for Magellanic Cloud members, black for other satellite galaxies) as their orbits in a Milky Way potential cannot be trusted. }
    \label{fig:dyn_group}
\end{figure}

Using the median apocentre and maximum height above the plane, calculated in Sect. \ref{sec:orbit_integration}, we quantified the number of stars belonging to different parts of the Galaxy. Based on these parameters, the upper panel of Figure \ref{fig:dyn_group} splits the sample into bulge (Apo$_{\rm mean}$ < 3.5 kpc, 3\%), inner halo (Z$_{\rm max}$ > 3.5 kpc, 3.5 kpc < Apo$_{\rm mean}$ < 20 kpc, 58 \%), outer halo (Z$_{\rm max}$ > 3.5 kpc, Apo$_{\rm mean}$ > 20 kpc, 28 \%), and confined stars (Z$_{\rm max}$ < 3.5 kpc, Apo$_{\rm mean}$ > 3.5, 11 \%). The stars are coloured by their L$_{\rm z}$ to distinguish between prograde and retrograde orbits.
As expected for VMP stars, and already illustrated by the distribution of Galactic latitudes in Figure \ref{fig:Skycov}, a majority of the VMP stars identified in this work are found in the halo. We do find a smaller subset of stars that seem to have more disc-like orbits with low eccentricities that are confined to the disc plane. While the smaller subset found to be confined in the disc could potentially include higher levels of contamination from the much more numerous more metal-rich (and younger) disc populations, we note that several studies have indeed found VMP stars confined to the disc \citep{Sestito2019, Sestito2020, DiMatteo2020A&A...636A.115D, FernandezAlvar2021MNRAS.508.1509F, FernandezAlvar2024A&A...685A.151F, Nepal2024A&A...688A.167N, PristineXXIV2024A&A...692A.131G, PristineXXV2025}. 

The lower panel of Figure \ref{fig:dyn_group} shows the distribution of energy versus the $z$ component of the angular momentum for the stars, with the stars coloured according to their orbital eccentricity. The distribution aligns well with the classifications made above. The polar orbits close to L$_{\rm z}=0$ kpc km s$^{-1}$ at moderate energy have higher eccentricities, consistent with being part of the halo. The mostly prograde confined stars at higher L$_{\rm z}$ have low eccentricities, most of them consistent with thick disc-like orbits, with 3 stars on lower Z$_{\rm max}$ matching the thin disc. There is evidence in the literature for a bias for prograde over retrograde orbits for metal-poor stars on similar disc-like orbits \citep{Sestito2020}.

In both panels of Figure \ref{fig:dyn_group} we have marked 34 stars that are probable members of dwarf galaxies with a black circle and a cross instead of a circle coloured by their properties. These stars will have internal velocities inside the dwarf galaxy, so the derived orbital properties in the (smooth) Milky Way potential will not be accurate. All stars marked here have probabilities > 0.99 that they belong to a dwarf galaxy based on the membership probabilities from \cite{Battaglia2022}. The overlap between our clean sample and \citet{Battaglia2022} contained one more star, but as its probability of belonging to a dwarf galaxy was 0, we have not included it in the coloured markers. 

Of the 34 dwarf galaxy stars, 7 do not have metallicity information from the sources in our literature comparison. Four of these belong to Ursa Minor, two to Sculptor and one to Sextans. The relative overabundance of Ursa Minor stars is likely due to high declination, making it more difficult to observe from the southern hemisphere. They do not appear in the catalogue of \cite{simbad2018AJ....156..257S}, or the recent large survey by \cite{Pace2020MNRAS.495.3022P}. All 7 of these stars are found in the outskirts of their dwarf galaxies.

A further 16 stars, identified in Figure \ref{fig:dyn_group} with special red crossed-circle markers, are identified as possible members of the Small or Large Magellanic Clouds. These stars have been observed as part of one of the LMC or SMC APOGEE fields and have proper motions and line-of-sight velocities that are in line with literature values for the Magellanic Clouds \citep{MC_PM2018}. 
The details of these stars are documented in Table tab\_c1, available at the CDS, as these stars will provide a useful addition to the limited number of known stars in the Magellanic Clouds at these low metallicities \citep{Reggiani2021AJ....162..229R, Oh2023Magellanic, Chiti2024NatAs...8..637C, Oh2024}. 

We note that one star of our Magellanic Cloud candidate stars overlaps with the study of \cite{Oh2023Magellanic}, as marked in the table with an o. In the high-resolution study from \citet{Oh2023Magellanic}, this star gets assigned a metallicity of [Fe/H]$= -2.62$, in line with what we expect for stars in our sample. A further 5 Magellanic Cloud VMP star candidates from \cite{Reggiani2021AJ....162..229R} have APOGEE spectra but are not included in our selection. As none of their literature metallicities are below $-2.5$, their exclusion from our sample is to be expected. None of the stars in the selection have photometric metallicities from \cite{PristineXXIII2023} or \cite{Andrae2023b} due to quality cuts on the CaHK magnitude and parallax respectively.

\subsection{Orbits for stars in the inner Galaxy}
\label{sec:dyn_res}

For stars in the inner Galaxy, we subdivided our sample into an `inner' and `outer' sample using R$_{\rm GC}$ and Apo$_{\rm mean}$, following the definition and nomenclature in \cite{ArdernArentsen2024}. Compared to the bulge definition from Section \ref{sec:distribution}, combining R$_{\rm GC}$ and Apo$_{\rm mean}$ allowed us to define both stars confined to the inner Galaxy, and halo interlopers that are temporarily in the inner regions. Both samples have $R_{\rm GC}$ < 3.5 kpc and are thus currently situated in the inner Galaxy in a region that is shared with the bulge. The first, inner, sample has median apocentre < 5 kpc, and the second, outer, sample has apocentres between 5 and 15 kpc. The full list of stars and their orbital properties is available at the CDS.

\begin{figure*}
    \centering
    \includegraphics[width=\linewidth]{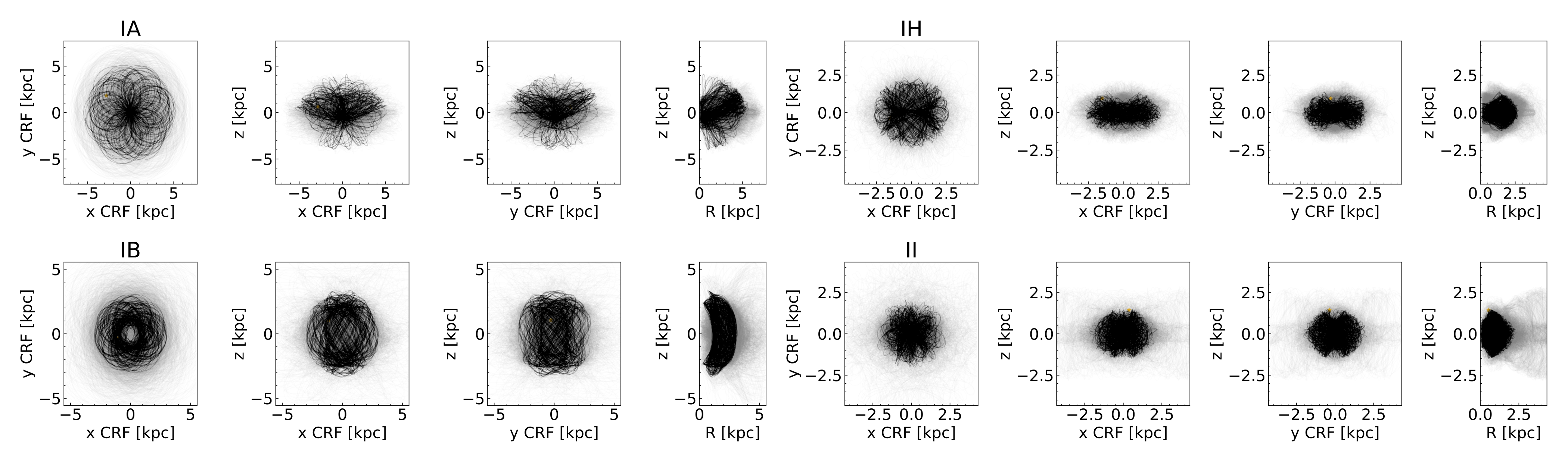}
    \caption{Orbits for stars within 3.5 kpc of the Galactic Centre and median apocentres within 5 kpc. The orbits are shown in a corotating frame, following the bar. The orbit is plotted in black, with 25 sampled orbits shown in grey. Each star will be identified with the two letter code shown above each row. The full plot is shown in the Appendix in Figure \ref{fig:bulge_orbits_inner}.}
    \label{fig:bulge_orbits_inner_cut}
\end{figure*}

Our inner sample consists of 14 stars, 8 of which have been observed previously by \cite{ArdernArentsen2024}. An additional 4 stars have existing photometric metallicities from \cite{PristineXXIII2023} or \cite{Andrae2023b}, but these stars have no spectroscopic derived parameters (and thus no orbits) in previous literature. A selection of the orbits for these stars are shown in Figure \ref{fig:bulge_orbits_inner_cut}; all relevant data for these stars are given in Table tab\_d1, available at the CDS\footnote{Stars observed by \cite{ArdernArentsen2024} are marked with a * in the table, stars observed by\cite{PristineXXIII2023} or \cite{Andrae2023b} are marked with a p.}. The full sample is available in Appendix \ref{app:orbs}. The orbits are shown in a corotating frame, following the rotation of the bar to highlight the influence of the bar on the stars. While the black line corresponds to the main orbit derived from the observed phase-space coordinates, an additional 25 orbits are derived from sampling the parameters within their uncertainties. They provide a view of the influence of the uncertainties on these orbits and are plotted in grey in each panel of Figure \ref{fig:bulge_orbits_inner_cut}. 

The outer sample has 13 stars. A selection of their orbits are shown in Figure \ref{fig:bulge_orbits_outer_cut}, with the full sample shown in Appendix \ref{app:orbs}, and all relevant data are given in Table tab\_d2 at the CDS. As indicated in the table, one star from this sample has spectroscopic data from PIGS in \cite{ArdernArentsen2024}. 
Additionally, 7 stars have previously derived photometric metallicities. 

As energy is not conserved in a time-dependant potential, we checked the Jacobian energy for all the stars in the inner Galaxy. For all these 27 stars the Jacobian energy is conserved up to an order of $10^{-6}$, ensuring the orbit integrations are robust. Due to the uncertainty in our distance estimates and the model dependence of the derived orbits, we caution for over-interpretation of the individual orbits. However, a general feature of these orbits is the high fraction of chaotic orbits. More specifically, $100\%$ of the inner sample orbits and $85\%$ of the outer sample orbits have Lyapunov exponents\footnote{The Lyapunov exponent is a measure of how chaotic an orbit is, with values higher than zero indicating a chaotic orbit, see \cite{LYAPUNOV01031992}.} greater than 0 for more than half of the 25 sampled orbits for the star. This ratio is significantly higher than what is usually found for more metal-rich parts of the bulge \citep[see e.g.][]{Valluri2016, Nieuwmunster2024A&A...685A..93N}. In part due to this chaoticity, none of the orbits meet the criteria of |y$_{\rm max}$|/|x$_{\rm max}$|$>0.35$ for a classical x1 orbit defined by \cite{Valluri2016} that are quite common among more metal-rich inner Galaxy stars, but with the relatively high uncertainties on our results, it is difficult to draw a firm conclusion.

The mean and standard deviation for the azimuthal ($v_\phi$), radial ($v_R$) and vertical ($v_z$) velocities are shown in Table \ref{tab:velocities}, along with upper and lower quantiles, as the low number of stars can make the dispersion unreliable. Despite using the same potential, comparing dynamical properties with the sample in \cite{ArdernArentsen2024} is complicated by low number statistics combined with the high standard deviation in the velocities of this kinematically hot population. Nevertheless, we note that our sample seems to have even less rotation in the mean compared to the results in \cite{ArdernArentsen2024}. This may be connected to the different selection as \cite{ArdernArentsen2024} had to avoid more dust-obscured regions for their selection of metal-poor stars using the optical $CaHK$ narrow-band filter or the different selection function in APOGEE, which prioritises bright stars.

\begin{table}
\caption{Means, upper and lower quantiles, and standard deviations of the azimuthal, radial, and vertical velocities for the inner and outer samples, as described in Sect. \ref{sec:dyn_res}.}           
\label{tab:velocities}      
\centering                         
\begin{tabular}{c c c c c}      
\hline\hline              
  \ & $\mu$ & q16 & q84 &  $\sigma$  \\  
  \ & [km/s] & [km/s] & [km/s] & [km/s] \\
\hline                       
   $v_{\rm \phi}$ Inner & 28 & -82 & 62 & 130 \\
   $v_{\rm \phi}$ Outer & -16 & -158 & 29 & 158 \\
   $v_{\rm R}$ Inner & 8 & -94 & 50 & 107 \\
   $v_{\rm R}$ Outer & 46 & -106 & 93 & 155\\
   $v_{\rm z}$ Inner & 31 & -131 & 89 & 138 \\
   $v_{\rm z}$ Outer & -15 & -168 & 14 & 155 \\
\hline                                 
\end{tabular}
\end{table}

\begin{figure*}
    \centering
    \includegraphics[width=\linewidth]{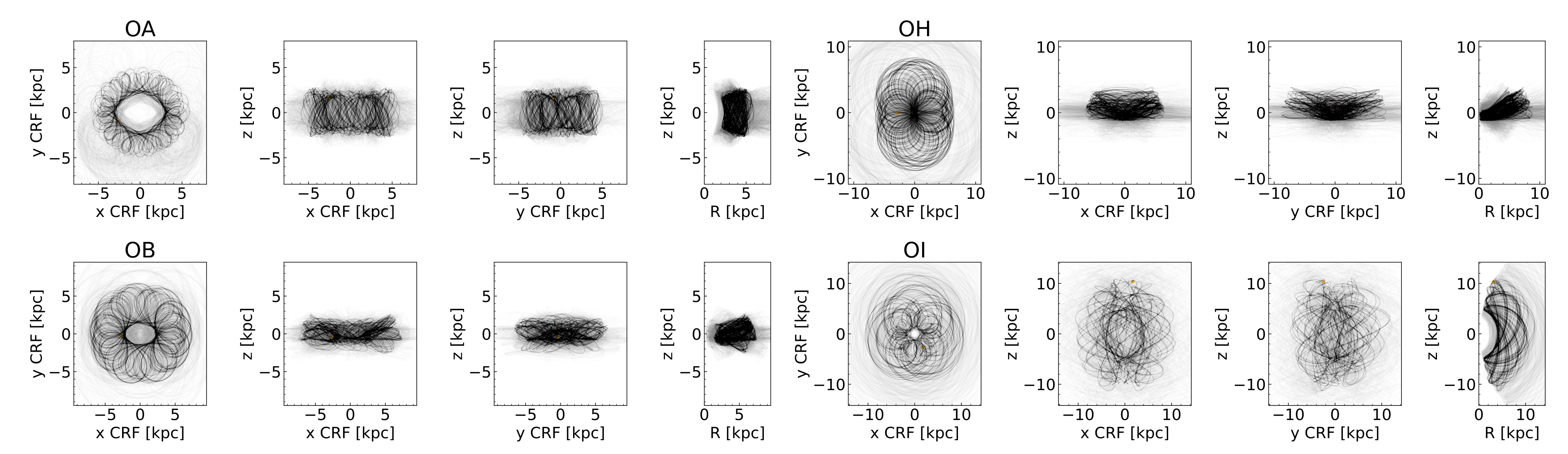}
    \caption{Orbits for stars within 3.5 kpc of the Galactic Centre and median apocentres between 5 and 15 kpc. The orbits are shown in a corotating frame, following the bar. The orbit is plotted in black, with 25 sampled orbits shown in grey. Each star will be identified with the two letter code shown above each row. The full plot is shown in the Appendix in Figure \ref{fig:bulge_orbits_outer}.}
    \label{fig:bulge_orbits_outer_cut}
\end{figure*}

\section{Discussion and conclusions}
\label{sec:summ}

In this paper we have studied how VMP stars are processed by the APOGEE survey`s analysis pipeline ASPCAP, and used that information to find a sample of candidate stars likely to have $\mh$ < $-2.5$. In particular, we have focused on stars with good quality spectra, no visible emission lines, and S/N $>$ 30, which nevertheless have gotten null values for their calibrated metallicities. Despite missing calibrated metallicities these stars have effective temperatures and surface gravities determined by the pipeline. Focusing on the stars with stellar parameters indicating they are members of the metal-poor red giant branch, yields a sample with robust comparisons to literature metallicities indicating a very high purity of stars with $\feh$ < -2. With these relatively simple cuts, we have isolated a sample of 289 (highly likely) VMP red giant branch stars, 53 of which lack metallicity measurements from the literature, and 83 which only have photometric metallicities. A comparison with spectroscopic literature samples shows no contamination of this sample with catastrophic (high-metallicity) outliers, while the success rate of this method to uncover stars below [Fe/H] $< -2.5$ is very high at 70\%. A smaller sample of 38 stars is not as well understood, with lacking literature data. 

Our results clearly show that it is possible to identify metal-poor stars with the existing information provided by the parameters provided by the APOGEE survey and minimal input from other sources. This means that for future surveys seeking to find metal-poor stars in dust obscured regions, the H-band is a viable option for such purpose, and for future H-band surveys \citep[such as MOONS,][]{MOONS2014} metal-poor stars can be identified with high confidence.

We note that our chosen selection criteria of APOGEE flags and parameters do not necessarily provide a unique pathway to identify these stars. Indeed, other methods of selecting metal-poor stars, such as selecting stars with low uncalibrated metallicities or with the $\metalsbad$ or $\metalswarn$ flags, have also been considered and tested for this work. A number of other flags are present for a large number of our stars, notably $\textsc{STAR\_WARN}$, $\textsc{SUSPECT\_ROTATION}$ and $\textsc{ROTATION\_WARN}$. While $\textsc{ROTATION\_WARN}$ appears to be most common among metal-poor stars, there is a significant number of metal-rich stars with this flag. No one flag seems to adequately single out metal-poor star candidates. 

While alternate methods can identify stars in the range $-2.5$ < $\mh$ < $-2.0$, very few additional stars with $\mh$ < $-2.5$ are added to the sample. For this reason, we have opted for the purpose of this study -- aimed at finding the most metal- poor stars -- to follow the method outlined in previous sections and selected stars based on missing $\apomh$ to reach higher purity.

By examining the orbital properties of the clean sample of stars we have identified, we have identified stars in regions of particular interest for spectroscopic follow-up. We found 27 stars in the inner Galaxy, with R$_{\rm GC}<3.5$ kpc. Of these 27, 14 have median apocentres confined within 5 kpc, with the remaining 13 having median apocentres between 5 and 15 kpc. The orbits appear highly chaotic, lacking classical x1 type orbits commonly seen at higher metallicities. 7 of these candidates have no literature metallicity, with an additional 9 only having photometric metallicities.
Out of 34 stars identified as part of a dwarf galaxy by \cite{Battaglia2022}, we found 7 stars without previous metallicity estimates, primarily in Ursa Minor. As the Magellanic Clouds are not part of \cite{Battaglia2022}, we have identified 16 potential members based on the APOGEE fields and line-of-sight velocities, Gaia proper motions, and the distances computed in this work. 15 of these candidates do not have metallicity estimates from the literature.

In paper II we analyse the spectra selected in this work, measuring magnesium and silicon abundances and use them to estimate overall metallicity for these stars. The full sample, including the lower confidence candidates, with all measurements, will be released with this paper. 
Additionally, we are currently performing follow-up studies of both interesting candidates from the sample, and of stars away from the RGB where other literature comparisons are scarce. 

\section*{Data availability}
Tables documenting our inner sample, outer sample, and Magellanic Cloud sample are available in electronic form at the CDS via anonymous ftp to cdsarc.u-strasbg.fr (130.79.128.5) or via http://cdsweb.u-strasbg.fr/cgi-bin/qcat?J/A+A/.

\begin{acknowledgements}
Thank you to the anonymous referee who provided swift and constructive feedback on this project. 
Thank you to Pascale Jablonka, Hanyuan Zhang and Danny Horta for helpful discussion and comments. 
M.M. and E.S. acknowledge funding through VIDI grant "Pushing Galactic Archaeology to its limits" VI.Vidi.193.093,  which is funded by the Dutch Research Council (NWO). This research has been partially funded from a Spinoza award by NWO (SPI 78-411). 
AV gratefully acknowledges support from the Canadian Institute for Theoretical Astrophysics (CITA) through a CITA National Fellowship and the International Astronomical Union (IAU) and the Gruber Foundation through a IAU Gruber Fellowship.
This research was supported by the International Space Science Institute (ISSI) in Bern, through ISSI International Team project 540 (The Early Milky Way). Guoshoujing Telescope (the Large Sky Area Multi-Object Fiber
Spectroscopic Telescope LAMOST) is a National Major Scientific
Project built by the Chinese Academy of Sciences. Funding for the
project has been provided by the National Development and Reform
Commission. LAMOST is operated and managed by the National
Astronomical Observatories, Chinese Academy of Sciences.
Funding for the Sloan Digital Sky Survey IV has been provided by the Alfred P. Sloan Foundation, the U.S. Department of Energy Office of Science, and the Participating Institutions. SDSS acknowledges support and resources from the Center for High-Performance Computing at the University of Utah. The SDSS web site is www.sdss4.org.
SDSS is managed by the Astrophysical Research Consortium for the Participating Institutions of the SDSS Collaboration including the Brazilian Participation Group, the Carnegie Institution for Science, Carnegie Mellon University, Center for Astrophysics | Harvard \& Smithsonian (CfA), the Chilean Participation Group, the French Participation Group, Instituto de Astrofísica de Canarias, The Johns Hopkins University, Kavli Institute for the Physics and Mathematics of the Universe (IPMU) / University of Tokyo, the Korean Participation Group, Lawrence Berkeley National Laboratory, Leibniz Institut für Astrophysik Potsdam (AIP), Max-Planck-Institut für Astronomie (MPIA Heidelberg), Max-Planck-Institut für Astrophysik (MPA Garching), Max-Planck-Institut für Extraterrestrische Physik (MPE), National Astronomical Observatories of China, New Mexico State University, New York University, University of Notre Dame, Observatório Nacional / MCTI, The Ohio State University, Pennsylvania State University, Shanghai Astronomical Observatory, United Kingdom Participation Group, Universidad Nacional Autónoma de México, University of Arizona, University of Colorado Boulder, University of Oxford, University of Portsmouth, University of Utah, University of Virginia, University of Washington, University of Wisconsin, Vanderbilt University, and Yale University. This research has made use of the SIMBAD database, operated at CDS, Strasbourg, France.
\end{acknowledgements}

\bibliographystyle{aa} 
\bibliography{references.bib}

\begin{appendix}

\section{Additional distance comparisons}
\label{app:dists}
\begin{figure*}
    \centering
    \includegraphics[width=0.8\linewidth]{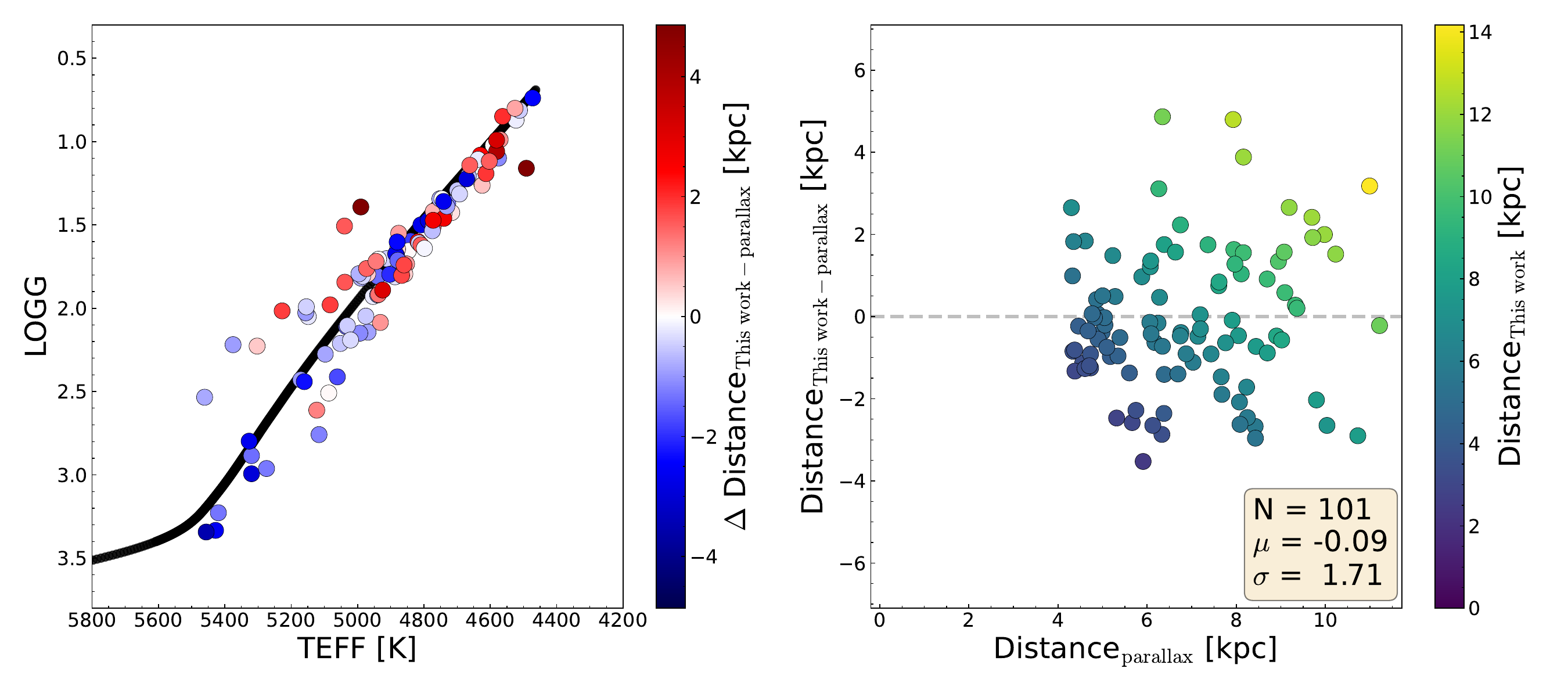}
    \caption{
    Left panel: Kiel diagram showing a BASTI isochrone corresponding to $\mh = -2.5$ and an age of 10 Gyr in black. Stars from the high confidence sample with $\frac{\pi}{\Delta\pi}>5$ to which distances have been determined are shown colour-coded by their offset from the distance determined by parallax. Right panel: Plot of the difference between our distances and the parallax distances plotted against parallax distance for the same stars as the left panel. }
    \label{fig:AV_dist_comp}
\end{figure*}

\begin{figure*}
    \centering
    \includegraphics[width=0.8\linewidth]{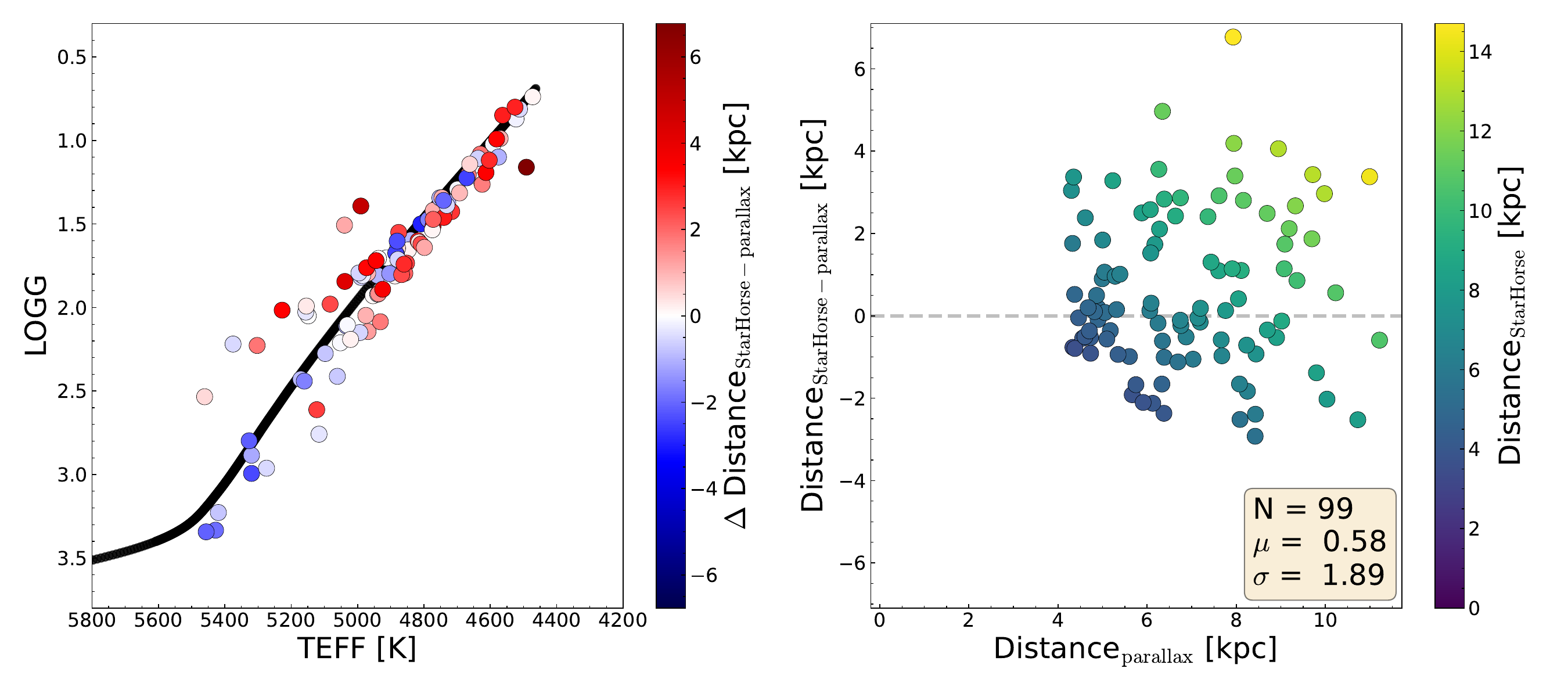}
    \caption{Same comparisons as in Figure \ref{fig:AV_dist_comp} but for the \textit{StarHorse} distances instead of our work. The reduction in sample size is due to stars lacking \textit{StarHorse} distances.}
    \label{fig:SH_dist_comp}
\end{figure*}

The validation of the distance determination performed in Section \ref{sec:dist} is shown in more detail in Figures \ref{fig:AV_dist_comp} and \ref{fig:SH_dist_comp} with a direct comparison of our distances and the ones from the \textit{StarHorse} \citep{Starhorse2023} catalogue which also incorporates spectroscopic data from APOGEE. The stars used to validate the distances are all taken from the high confidence sample. A quality cut of $\frac{\pi}{\Delta\pi}>5$ is applied to ensure a reliable distance can be determined from the parallax alone. Additionally, only stars with parallax distances in the range $4.3$ kpc $< r_{\pi}< 11.3$ kpc are included, to mimic the distance range of the inner Galaxy which we are aiming to study in more detail. Note that there is no cut on the sky, so the stars will be distributed over the whole sky and not confined to the inner Galaxy.
The final sample ends up with 100 stars out of the 297 in the high confidence sample.

\begin{figure}
    \centering
    \includegraphics[width=0.8\linewidth]{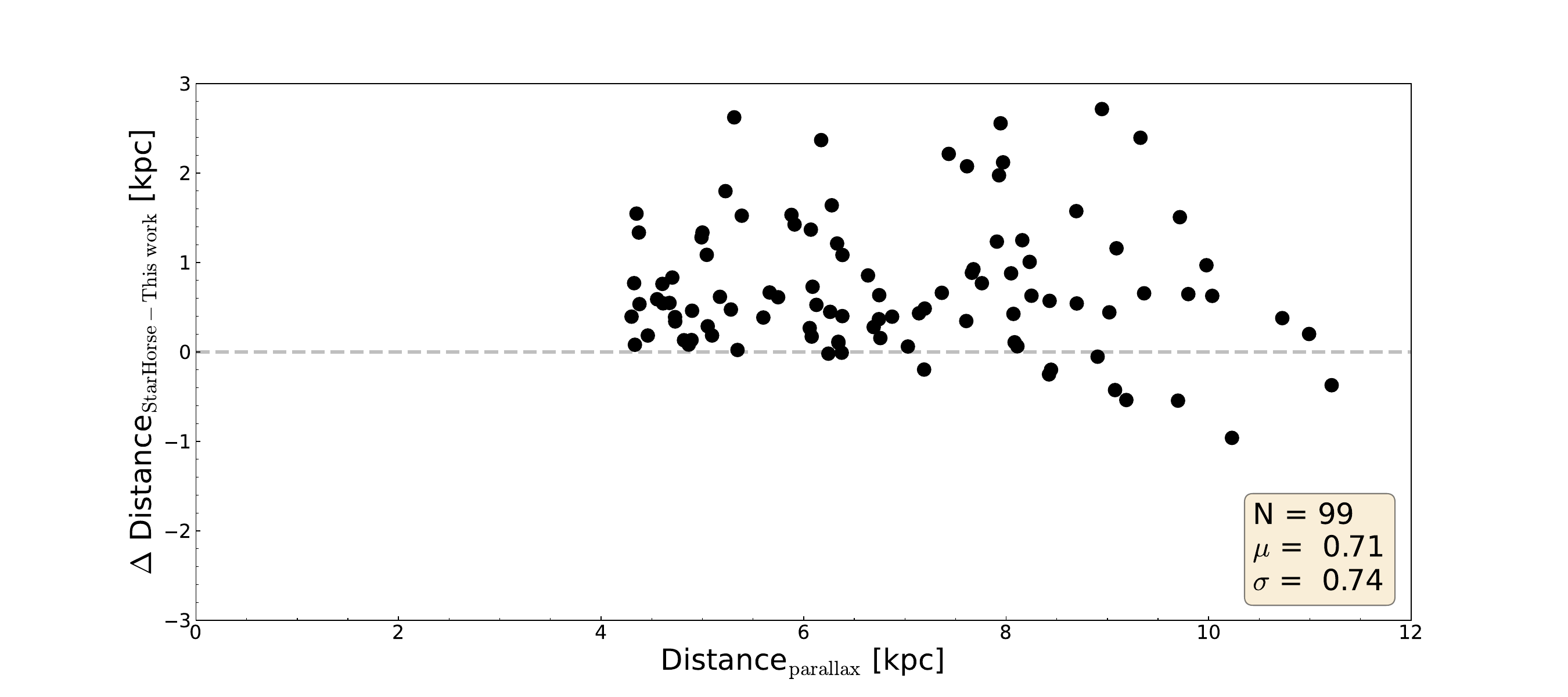}
    \caption{Direct comparison of our work and \textit{StarHorse} for the stars presented in figures \ref{fig:AV_dist_comp} and \ref{fig:SH_dist_comp}.}
    \label{fig:AV_SH_dist_comp}
\end{figure}

For both Figure \ref{fig:AV_dist_comp} and \ref{fig:SH_dist_comp} the left panels show a Kiel diagram with a BASTI isochrone corresponding to $\mh = -2.5$ and an age of 10 Gyr in black and the difference between our results (or \textit{StarHorse} Dist50 column) and the parallax distance, calculated by inverting parallax, is shown in a blue-red colour scale. The colourbar for the left panels has been normalised to the highest error in both panels, centred on 0, for easier comparison. The right panel shows the comparison of our (\textit{StarHorse's}) distance and parallax distance used as the colourbar in the left panel, but here plotted against the parallax distance and coloured by our (\textit{StarHorse's}) distance. Our distances and the ones from \textit{StarHorse} are directly compared against the parallax distance in figure \ref{fig:AV_SH_dist_comp}.

Figures \ref{fig:AV_dist_comp} and \ref{fig:SH_dist_comp} show a remarkably similar distribution in both panels. As reported in Section \ref{sec:dist} the mean offset and standard deviation of our distances are somewhat lower. The left panels attempt to find a correlation between the offset from parallax distance and position in the Kiel diagram compared to the BASTI isochrone used to measure our distances. The clearest relation is found for the lowest end of the giant branch, where the distances to the subgiants are overestimated by both our and \textit{StarHorse's} distances. The turn-off is the point where the greatest discrepancies between isochrones and stellar parameters are seen, both in this work and in the literature \cite[see e.g.][]{daSilva2023A&A...677A..74D}. However, as the stellar parameters for these stars appear well aligned with the isochrone, the reason for this discrepancy is not clear.

Another commonly used set of distances are the ones from \cite{2021AJ....161..147B}. Due to how these distances are determined and the nature of our sample, we feel that a similar comparison with the Bailer-Jones distances could be misleading. As the Bailer-Jones distances rely on parallax when the parallax quality is high (and more heavily on a prior when the parallax becomes unreliable), comparing the Bailer-Jones distances against good parallax stars will compare favourably because both axes are far from independent. As $\sim$1/2 of the entire high confidence sample (not just the validation sample) does not meet the requirement of $\frac{\pi}{\Delta\pi}>5$, it will not be a fair representation of the whole sample. For the poorer parallax half of the sample, the Bailer-Jones distances will rely more on the assumed prior of the distribution of the stars in the Galaxy. Since the stars we are most focused on are in the inner Galaxy, the main prior will come from the more massive, more metal-rich part of the bulge. In contrast, we are studying a trace population of the most metal-poor stars and we prefer to avoid being biased by the bulk of the bulge stars. For these reasons, we believe that a method that is independent of the quality of parallax and does not rely on a density distribution prior (such as the isochrone method we apply) is much better suited to our science case.

\section{Inner galaxy orbits}\label{app:orbs}
\begin{figure*}
    \centering
    \includegraphics[width=\linewidth]{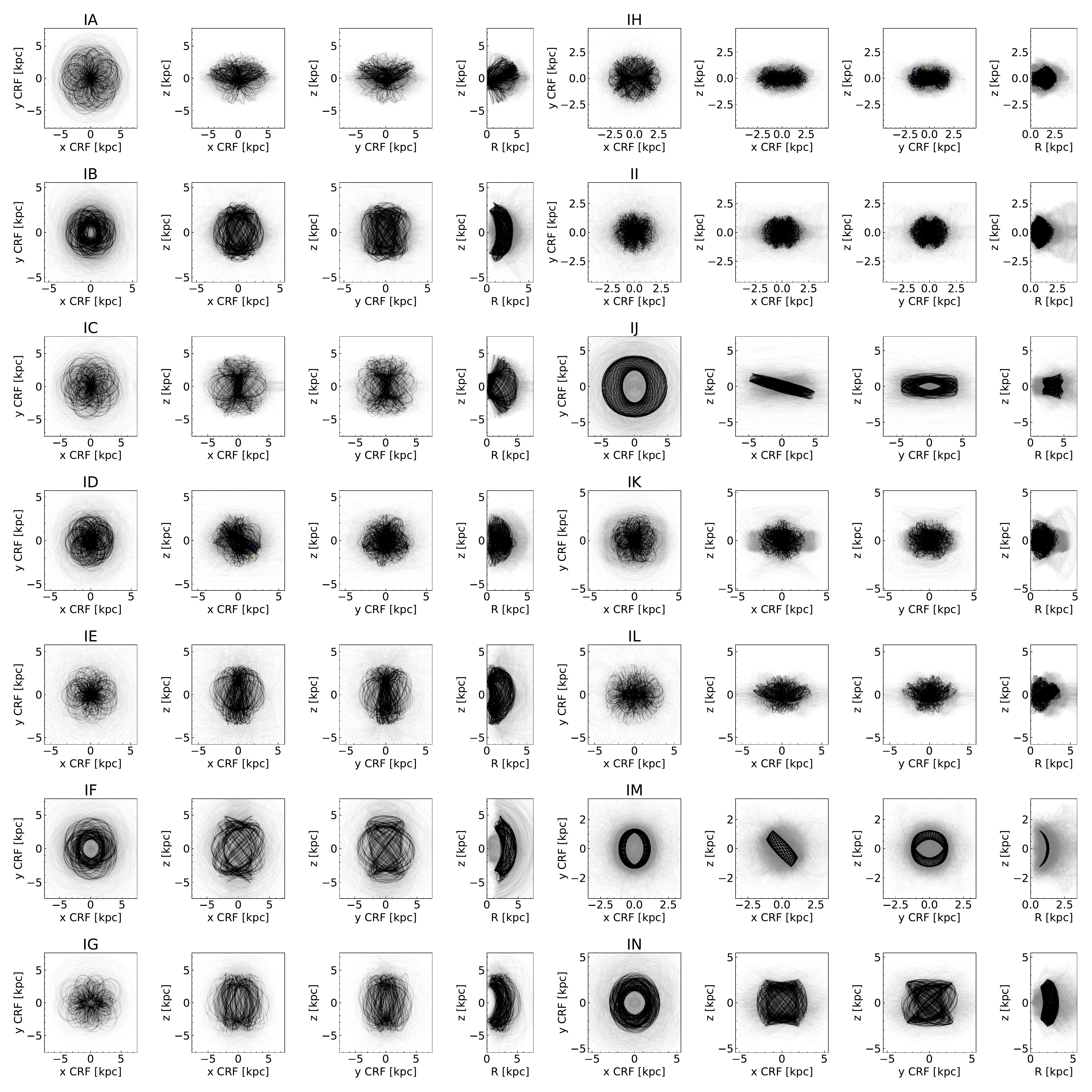}
    \caption{Orbits for stars within 3.5 kpc of the Galactic Centre and median apocentres within 5 kpc. The orbits are shown in a corotating frame, following the bar. The orbit is plotted in black, with 25 sampled orbits shown in grey. }
    \label{fig:bulge_orbits_inner}
\end{figure*}

\begin{figure*}
    \centering
    \includegraphics[width=\linewidth]{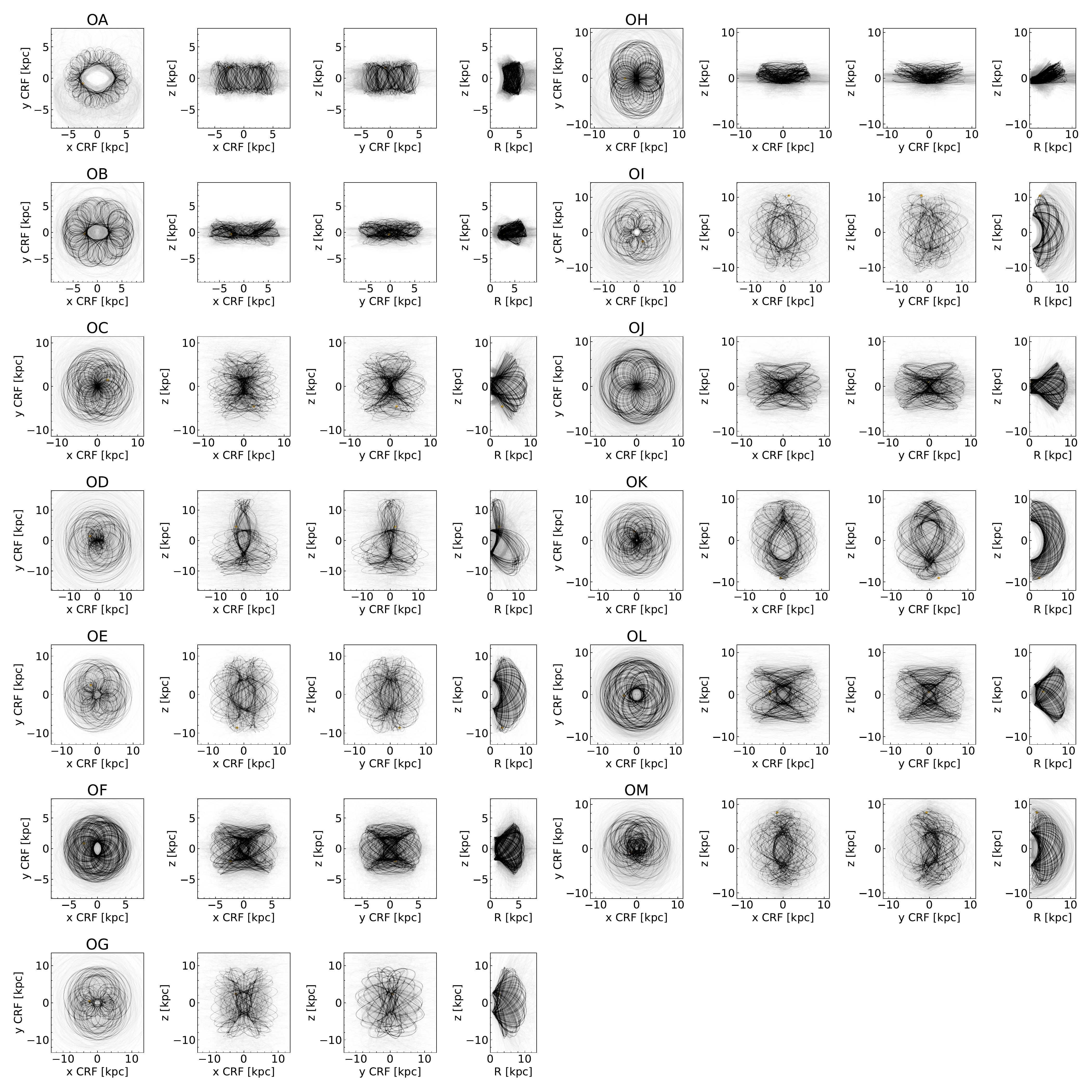}
    \caption{Orbits for stars within 3.5 kpc of the Galactic Centre and median apocentres between 5 and 15 kpc. The orbits are shown in a corotating frame, following the bar. The orbit is plotted in black, with 25 sampled orbits shown in grey. }
    \label{fig:bulge_orbits_outer}
\end{figure*}

Figures \ref{fig:bulge_orbits_inner} and \ref{fig:bulge_orbits_outer} show the complete versions of Figures \ref{fig:bulge_orbits_inner_cut} and \ref{fig:bulge_orbits_outer_cut}, where only the first two rows are shown. OI and OK in Figure \ref{fig:bulge_orbits_outer} are the stars with a lower degree of chaoticity, with $48\%$ and $32\%$ of sampled orbits being chaotic respectively.

\end{appendix}

\end{document}